\documentclass[12pt,a4paper]{article}
\usepackage{geometry}
\usepackage[round]{natbib}
\usepackage{mathtools}
\allowdisplaybreaks
\usepackage{mathrsfs}
\usepackage{amsfonts}
\usepackage{amssymb}
\usepackage{graphicx}
\usepackage{placeins}
\usepackage{stmaryrd}
\usepackage{lipsum}  
\usepackage[labelfont=bf]{caption}
\usepackage[skip=-0.25\textwidth]{subcaption}
\usepackage{enumitem}
\usepackage{titlesec}

\titleformat{\section}
{\normalfont\fontsize{12}{15}\bfseries}{\thesection .}{0.3em}{}
\titleformat{\subsection}
{\normalfont\fontsize{12}{15}\itshape\bfseries\centering}{\upshape\thesubsection .}{0.3em}{}
\titleformat{\subsubsection}
{\normalfont\fontsize{12}{15}\itshape\bfseries}{\upshape\thesubsubsection .}{0.3em}{}

\usepackage{color}
\usepackage[titletoc,title]{appendix}
\usepackage{hyperref}%
\hypersetup{
	colorlinks   = true,
	citecolor    = blue
}
\usepackage{xpatch}
\usepackage{placeins}
\renewenvironment{abstract}  
{\normalfont
	\begin{center}
		\bfseries \abstractname\vspace{-.5em}\vspace{0pt}
	\end{center}
	\list{}{%
		\setlength{\leftmargin}{0.5 mm}
		\setlength{\rightmargin}{\leftmargin}%
	}%
	\item\relax}
{\endlist}  
\def\equationautorefname~#1\null{%
	(#1)\null
}   
\xpretocmd{\eqref}{equation\,}{}{}
\usepackage{authblk}

\title{\textbf{Near-wall hydrodynamic slip triggers swimming state transition of microorganisms}}
\author[1]{Antarip Poddar}
\author[1]{Aditya Bandopadhyay\thanks{Email: aditya@mech.iitkgp.ernet.in}}
\author[1]{Suman Chakraborty\thanks{Email: suman@mech.iitkgp.ernet.in}}
\affil[1]{Department of Mechanical Engineering, Indian Institute of Technology Kharagpur, Kharagpur, West Bengal - 721302, India}
\date{\normalfont\today}
\begin{document}
			\maketitle
\linespread{1.1}\begin{abstract}
\noindent
Interaction of motile microrganisms with a nearby solid substrate is a well studied phenomenon. However, the effects of hydrodynamic slippage on the substrate have received a little attention. In the present study, within the framework of the squirmer model, we impose a tangential velocity at the swimmer surface as a representation of the ciliatory propulsion and subsequently obtain exact solution of the Stokes equation based on a combined analytical-numerical approach. We illustrate how the near-wall swimming velocities are non-trivially altered by the interaction of wall slip and hydrodynamic forces. We report a characteristic transition of swimming trajectories for both puller and pusher type microswimmers by hydrodynamic slippage if the wall-slip length crosses a critical value.  In case of puller microswimmers that are propelled by a breast-stroke like action of their swimming apparatus ahead of their cell body, the wall slip can cause wall-bound trapping swimming states, either as periodic or damped periodic oscillations which would otherwise escape from a no slip wall. The associated critical slip length has a non-monotonic dependence on the initial orientation of the swimmer which is represented by novel phase diagrams. Pushers, which get their propulsive thrust from posterior flagellar action, also show similar swimming state transitions but in this case the wall slip mediated reorientation dynamics and the swimming modes compete in a different fashion to that of the pullers. Although neutral swimmers lack a sufficient reorientation torque to exhibit any wall-bound trajectory, their detention time near the substrate can be significantly increased by tailoring the extent of hydrodynamic slippage at the nearby wall. The present results pave the way for understanding the motion characteristic of biological microswimmers near confinements with hydrophobic walls or strategize the design of microfluidic devices used for sorting and motion rectification of artificial swimmers by tailoring their surface wettability.

	\end{abstract}

\section{Introduction}
Microswimmers encountering a confining geometry are common occurrences in a plethora of biological scenarios such as marine ecosystem, animal body as well as in controlled microfluidic lab-on-a-chip devices \citep{Bechinger2016,Denissenko2012}. One of the most important practical applications of the surface-microorganism interaction is the bacterial entrapment near surfaces, which is regarded as an essential step during biofilm formation \citep{Costerton1987}. In addition, a confining surface has been found to cause a host of intriguing phenomena ranging from directional circular motion of motile cells near a solid surface or an air-liquid interface \citep{Lauga2006,Lemelle2010,DiLeonardo2011}, scattering of \textit{Chlamydomonas} algae cells \citep{Molaei2014}, suppression of the tumbling motion of bacteria \textit{Escherichia coli} \citep{Kantsler2013} to pairwise dancing of \textit{Volvox} \citep{Drescher2011} etc. Such elemental near-surface behviour of motile cells is found to affect various biophysical activities such as guidance of sperm cells through female oviduct \citep{Guidobaldi2015,Ishimoto2015} and also crucially affects the process of  bacterial infection \citep{Harkes1992}. With recent advancement of microfluidics techniques, different artificial microswimmers have been successfully fabricated with promising applications  ranging from biochemical sensing, targeted drug delivery to environmental remediation \citep{Duan2015,Campuzano2017,Poddar2019,Richard2018}. The interfacial properties of the microfluidic chips \citep{Das2015,Simmchen2016} can be exploited to gain a control over the design of such synthetic microswimmers. In a recent experimental study, researchers\citep{Ketzetzi2018} observed an enhanced swimming speed of spherical self-diffusiophoretic swimmers near hydrophobic substrates. 

Inspired by their fascinating trends of near-surface swimming, different theoretical models of microswimmers have been proposed to physically describe their kinematics of motion  \citep{Berke2008,Zargar2009,Or2009,Crowdy2011,Spagnolie2012,Spagnolie2015,Ishimoto2013,Li2014,Shum2010,Mathijssen2016a,Daddi-Moussa-Ider2018,Walker2019,Kuron2019,Desai2018}. 
Employing a force-dipole swimmer model, \citep{Berke2008} was able to explain the high concentration of bacteria \textit{Escherichia coli} near glass surfaces as a consequence of hydrodynamic attraction and wall-parallel reorientation of the swimmer by its image. 
Motile microorganisms such as \textit{Opalina,Volvox} and \textit{Paramecium} have been widely modeled by considering a deformable spherical cell body with external appendages such as cilia or flagella on them performing  small-amplitude periodic beating and causing a bulk streaming of their cell surface. In the absence of inertial effects, these organisms show a force-free swimming \citep{Lauga2009}. These model microswimmers, popularly known as `squirmers', \citep{Lighthill1952,Blake1971} have been used to understand a variety of physical phenomena which include but are not limited to hydrodynamic interaction of two microswimmers \citep{Ishikawa2006}, diffusion and suspension rheology \citep{Ishikawa2007}, nutrient uptake \citep{Magar2005}, rheotaxis \citep{Uspal2015} and density stratification of suspending medium on the vertical motion of microswimmer \citep{Doostmohammadi2012}. Spherical squirmers and its variants have also been used to analyze the microswimmer behaviour near confinements \citep{Spagnolie2012,Ishimoto2013,Li2014,Yazdi2017}.

The wettability  of the confining substrate can severely influence the near-wall flow and the interfacial friction of the fluid, leading to interesting consequences at the micro and nanoscale \citep{Chakraborty2008,Maduar2015}. This is characterized by slip length, defined as the extrapolation distance below the surface where tangential fluid velocity would vanish. Hydrophilic surfaces, in contact with aqueous solutions, give rise to a negligible hydrodynamic slippage while the slip length lies in the range of a few tens of nanometers for smooth hydrophobic surfaces \citep{Huang2008,Bocquet2010}. On the other hand, specially treated nanostructured surfaces  or the presence of depleted, low viscosity,  wall-adjacent regions in bacterial polymeric solutions  often lead to an  augmented partial slip with the slip length ranging in micrometers  \citep{Joseph2006,Lee2008,Zhu2001,Tretheway2002,Lauga2007,Kaynan2017}.

Most of the previous research studies related to the locomotion of microswimmers near confinements were based on the no-slip walls or air-liquid interface characterized by an infinite fluid slip. The consequences of a partial slip boundary have received less attention in the past \citep{Lopez2014,Hu2015,Lemelle2013}.
In their experimental investigation, \cite{Lemelle2013} observed a reversal of circular wall-parallel trajectories of \textit{E.Coli} with addition of polymeric inclusions in the swimming medium and attributed the phenomenon as an effect of enhanced slip. Subsequently, the results of the mesoscopic simulations of \cite{Hu2015} showed a similar shift of clockwise to anticlockwise rotation of \textit{E. Coli} in a plane parallel to the surface. In addition they showed that a patterned surface with different slip lengths can be used to direct bacterial motion.
The far-field analysis of \cite{Lopez2014} employed an image singularity solution applied to a force dipole swimmer which predicted that a partial slip condition at the nearby surface will impart a wall-faced rotation and will always attract a pusher type swimmer. Even within the far-field analysis, they did not consider the contributions from the higher order singularities arising from a finite size cell body or the fore-aft asymmetry of the swimmer, which were found to have profound effect on the motion characteristics near a no-slip wall \citep{Spagnolie2012}. 
Also, an overestimation of the near-field hydrodynamic interactions by the far-field analysis \citep{Lopez2014} was revealed in the numerical simulations of \cite{Hu2015}.

The orientation dynamics of a microswimmer taking place in close proximity to a wall has been reported to exhibit diverse trajectory characteristics, ranging from wall escape to wall-induced stable trapping \citep{Li2014,Ishimoto2013,Ishimoto2017,Lintuvuori2016}. Beyond a far-field prediction based on fundamental singularities of Stokes flow, a  more detail account of the near wall hydrodynamic effects is necessary to explore the resulting trajectory as the microswimmer approaches a wall  \citep{Ishimoto2013,Li2014,Bechinger2016}.  In the present work,  we employ the squirmer model for spherical cell bodied swimmers, and within the realm of Stokes flow we obtain exact solution of the governing equations under interfacial slip, by exploiting a combined analytical-numerical method based on eigenfunction expansion in bispherical coordinates. The results indicate that wall slip beyond a stipulated strength can cause intense characteristic modifications in the swimmer trajectories of different types of microswimmer. 
	In this noteworthy that periodic and damped oscillatory trajectories, have been reported by previous numerical simulations \citep{Lintuvuori2016,Ishimoto2017} where the wall is  repulsive in nature, albeit with no hydrodynamic slippage. In sharp contrast, the exclusiveness of the present study lies in identifying and characterizing different swimming states in the presence of wall slip and how the enhancement in slip modulates different swimming aspects observed near a no slip wall. 	

\section{Mathematical description}
\label{sec:Mathematical}

\subsection{Problem formulation}
\label{ssec:formulation}

\begin{figure}[!htb]
	\centering
	\includegraphics[width=0.75\textwidth]{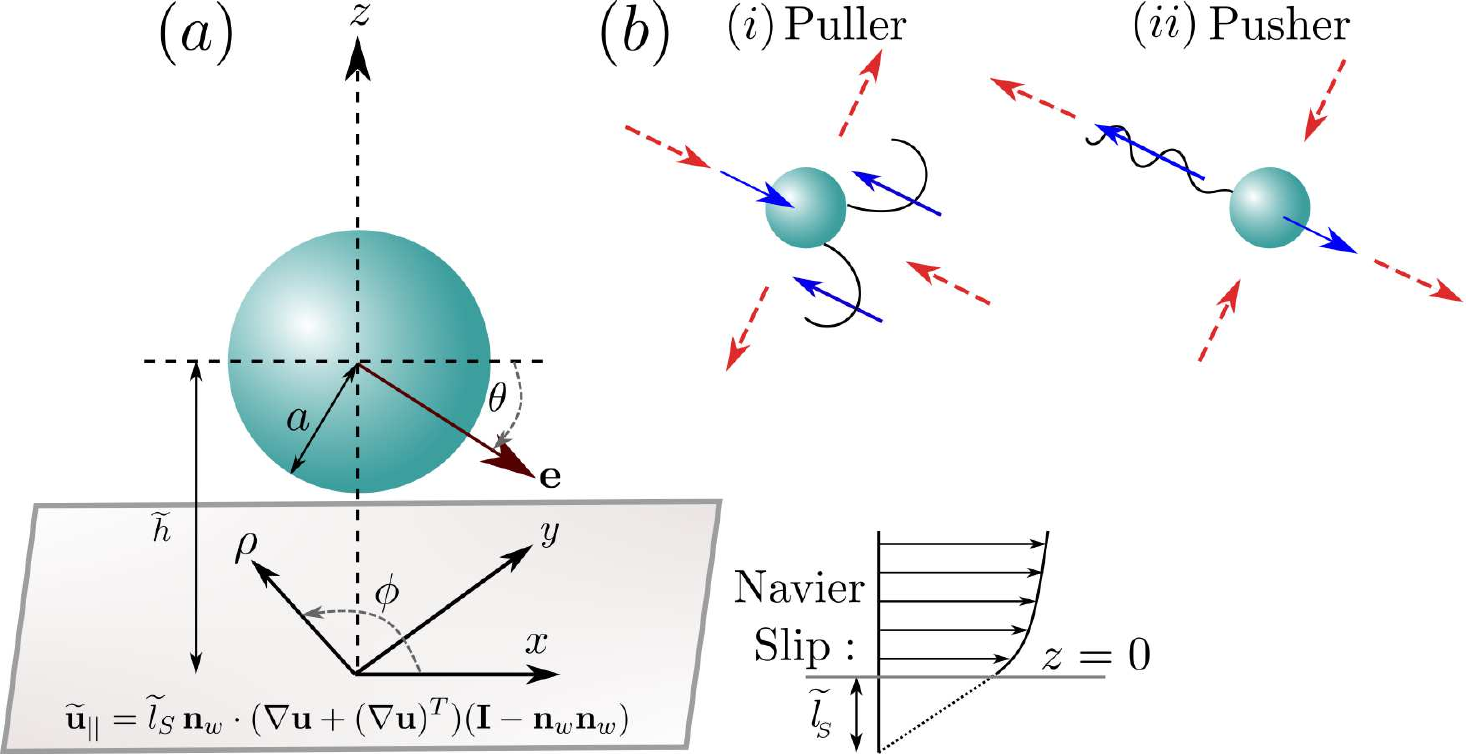}
	\vspace{1.5ex}
	\caption{Schematic representation of a model microswimmer near a slippery flat surface obeying Navier slip condition. Subfigure (a): The microswimmer has a spherical cell body with radius $ a $. The direction of swimmer thrust or the director vector is indicated by $ \mathbf{e}$, which is oriented at an angle of $ \theta $ from the positive $ x $ axis. The adjacent flat surface has a slip length of $ \widetilde{l}_S.$ Subfigure (b): Puller and pusher swimmers having two different propulsion mechanisms are schematically shown in an unbounded domain. Red dashed arrows show surrounding fluid flow while blue arrows indicate local forcing directions of the microswimmer to the fluid when viewed from the lab frame.} 
	\label{fig:schematic}
\end{figure}

We consider the quasi-steady motion of a microswimmer in a Newtonian fluid near a solid surface, along which the no-slip condition of fluid velocity is violated and hydrodynamic slippage takes place. The schematic description of the problem geometry is presented in figure~\ref{fig:schematic}(a). The spherical cell body of the model microswimmer has a radius $ a$ and its center is at a distance $ \widetilde{h} $ from the adjacent slippery wall. The size of the microswimmer is enough small to neglect the inertial effects and at the same time not small enough so that the Brownian effects become dominant. The swimmer thrust is along $ \mathbf{e}$, which is at a polar angle of $ \theta $ from the positive $x$-axis that is fixed at the wall and at an azimuthal angle $ \phi $ from the $ x-z$ plane. The counterclockwise rotation along the $ y $ axis is taken as positive. Here $ \widetilde{l}_S$ is the slip length denoting the extrapolation distance below the surface where the tangential fluid velocity would vanish.

Neglecting the inertial effects, the flow field around the swimmer can be described by the incompressibility condition and Stokes equation as 
\begin{equation}\label{eq:creeping}
 \nabla \cdot \widetilde{\mathbf{v}}=0     \quad \text{and} \quad -\nabla \widetilde{p}+\mu \nabla ^2 \widetilde{\mathbf{v}}=0.
\end{equation}
The hydrodynamic slippage at the confining wall is characterized by the Navier slip boundary condition \citep{Navier1823}, where the surface slip velocity has a linear variation with the shear rate at the plane surface, given as
\begin{equation}\label{eq:Navier_slip}
\widetilde{\mathbf{u}}_{||}=\widetilde{l}_S \, \mathbf{n}_w \cdot  (\nabla \mathbf{u}+(\nabla \mathbf{u})^T ) (\mathbf{I}-\mathbf{n}_w \mathbf{n}_w),
\end{equation}
where $ \mathbf{n}_w $ is the unit normal at the 	plane wall pointing into the fluid and $\mathbf{u}_{||}$ is the velocity component tangent to the plane wall. 
The microswimmer gains motility from the surface distortions generated by their swimming appendages. Following the `squirmer' model by Lighthill and Blake \citep{Lighthill1952,Blake1971} we impose a tangential velocity on a particle surface which mimics the locomotion of microbes due to ciliatory beating on their surface. The tangential surface velocity of a squirmer has the form 
\begin{equation}
\label{eq:tangential_vel_general}
\widetilde{\mathbf{u}}^s=\left(\frac{\mathbf{e}\cdot \mathbf{r} }{|\mathbf{r}|}\frac{\mathbf{r}}{|\mathbf{r}|}-\mathbf{e}\right) \sum_{n=1}^{\infty} \frac{2}{n(n+1)} B_n P'_n\left(\frac{\mathbf{e}\cdot \mathbf{r} }{|\mathbf{r}|}\right),
\end{equation}
where $ \mathbf{e} $ is the orientation vector of the director of the swimmer, $\mathbf{r}$ is the position vector of an arbitrary point on the swimmer surface with respect to the particle center, $B_n$ denotes the $n$-th squirming mode amplitude and $P'_n$ is the derivative of the Legendre polynomial, $ P_n$. The tangential velocity is assumed to be time-independent and represents an average over numerous beating cycles. Following earlier studies \citep{Ishikawa2006,Li2014,Shen2018,Yazdi2017,Shaik2017}, we consider only the first two squirming modes. 
Depending on the ratio of the first two squirming mode amplitudes, we define a squirmer parameter, $\beta=B_2/B_1$, which characterizes the intensity of the stresslet exerted by the swimmer. Pusher type swimmers, e.g. bacteria or sperm cells, which have the flagella behind the main cell body, correspond to $ \beta>0$, while in contrast, pullers have their flagella in the front, e.g. Chlamydomonas. The distinct propulsion mechanisms of these swimmers are shown schematically in figure~\ref{fig:schematic} (b). 
Also, $ \beta=0 $ denotes the class of swimmers generating symmetric flow field and are designated as Neutral swimmers, e.g. Volvox. We non-dimensionalize the lengths by the swimmer radius $ a $, velocity by  $U_{ref}=2 B_1 /3  $ (so that the unbounded medium squirming velocity becomes unity), time by $ a/U_{ref} $ and pressure by $ \mu U_{ref}/a$. Hereafter, the normalized variables will be denoted without the `$\; \widetilde{}\;$' symbol.
 If the microswimmer has a translational velocity of $ \mathbf{V}$ and a rotational velocity of  $\boldsymbol{ \Omega} $, then in the laboratory frame, the boundary condition for the fluid velocity at the surface of the swimmer can be written as 
 \begin{equation}\label{eq:BC_swimmer}
 \mathbf{u}_s=\mathbf{V}+\boldsymbol{\Omega} \times \mathbf{r} + \mathbf{u}^s.
\end{equation}
 In addition, since the swimmer is assumed to be neutrally buoyant in the suspending fluid, it experiences zero net force and zero net torque about its center, i.e.
 \begin{equation}\label{eq:force_free}
 \mathbf{F}= \iint \limits_{S_p}^{} \boldsymbol{\sigma}\cdot \mathbf{n_p}\, dS=0 \quad \text{and} \quad \mathbf{L}= \iint \limits_{S_p}^{} \mathbf{r} \times (\boldsymbol{\sigma}\cdot \mathbf{n_p})\, dS=0,
  \end{equation}
where $ \boldsymbol{\sigma} $ is the stress tensor and $ \mathbf{n_p} $ is the unit outward normal to the swimmer surface $ S_p$. Now, solving \eqref{eq:creeping} along with the boundary conditions
\eqref{eq:Navier_slip} and \eqref{eq:BC_swimmer}, one can obtain $\mathbf{V}$ and $ \boldsymbol{\Omega} $ by satisfying  \eqref{eq:force_free}.

Owing to the axisymmetric squirmer surface velocity in the present model \eqref{eq:tangential_vel_general}, the director $ \mathbf{e} $ of the swimmer is confined in the $ x-z $ plane. It will rotate around an axis which is perpendicular to both the wall-normal and body-fixed normal $ \mathbf{n_p}$ i.e. along the direction $ \mathbf{n_w} \times \mathbf{n_p}$. For the chosen co-ordinate system, this lies along the $ y $ axis. Hence, the locomotion of the microswimmer can be described by $ \{ \mathbf{V},\boldsymbol{\Omega} \}=\{V_x \mathbf{e}_x+V_z \mathbf{e}_z,\Omega_y \mathbf{e}_y\}.$ 

\subsection{Exact solution using Bispherical coordinates}
\label{ssec:bisp}
The Stokes equation, coupled with the pertinent boundary conditions including interfacial slip, is solved in terms of the eigensolutions for bispherical coordinates $ (\xi, \eta, \phi).$ The velocity components are evaluated in a cylindrical coordinate system $ (\rho,z,\phi) $  having its origin at the plane wall, the z-axis being normal to the wall and passing through the center of the spherical swimmer body. The bispherical and cylindrical coordinates are related as \citep{Happel2012}
\begin{equation}\label{eq:bisp_def}
\rho=c\frac{\sin(\eta)}{\cosh(\xi)-\cos(\eta)} \quad \text{and} \quad z=c\frac{\sinh(\xi)}{\cosh(\xi)-\cos(\eta)}, 
\end{equation}
where $ c $ is a positive scale factor. Here $ \xi=0 $ represents the plane wall and $ \xi=\xi_0 $ (where $ \xi_0 >0 $) denotes the surface of the sphere which has its center at $ z=c \, \coth(\xi_0) $ and has a radius of $ c/\sinh(\xi_0)$.
The general solution of the flow field was given by \cite{Lee1980} with the help of 7 unknown constants ($ A_n^m,B_n^m, C_n^m, E^m_n, F_n^m, G_n^m, H^n_m $) and associated Legendre polynomial ($ P_n^m=P_n^m(\cos(\eta)) $). Using the same general solution, researchers have solved the flow fields due to  squirming microswimmer problems near a two fluid interface \citep{Yazdi2017,Shaik2017} or the problem of a diffusiophoretic swimmer near a no-slip plane wall \citep{Mozaffari2016}. In sharp contrast, here, the situation is more complex since both the squirmer boundary condition at the swimmer surface (\eqref{eq:BC_swimmer}) as well as the Navier slip condition at the plane wall (\eqref{eq:Navier_slip}) are to be satisfied while obtaining the arbitrary constants. 
In the cylindrical coordinates, the slip boundary condition at the plane wall reads
\begin{equation}\label{eq:slip_cyl}
u_\rho=l_{\!_S} \, \sigma_{\rho z} \qquad \text{and} \qquad u_\phi=l_{\!_S} \, \sigma_{\phi z} \quad \text{at}
\; z=0.
\end{equation} 
Using the no-penetration of fluid at this surface, the above equation gets simplified to 
 \begin{equation}\label{eq:simplify_slip}
 u_\rho=l_{\!_S} \frac{\partial u_\rho}{\partial z} \quad \text{and} \quad  u_\phi=l_{\!_S} \frac{\partial u_\phi}{\partial z} \quad \text{at} \; z=0.
 \end{equation}
It is noteworthy to observe that although acting in a regime of low Reynolds number flow, the wall slip effects are not obtained as a trivial extension to previously researched studies on a microswimmer near a no-slip wall. Also, the present approach differs from the  asymptotic perturbation approach in terms of a small slip length as a perturbation parameter, which has been employed previously for unbounded particles with inhomogeneous surface slip \citep{Swan2008,Ramachandran2009,Willmott2008}. In effect, our results demonstrate that the effects of the fluid slip at the wall, manifested through the dimensionless slip length $ l_{\!_S} $, modify the velocity field in a rather intriguing and non-trivial manner. Further details regarding the solution procedure have been provided in  \ref{sec:sol_detail}. 
The exact solution approach deployed by us, using bispherical coordinates, can incorporate any separation distance from the wall and any degree of wall slip \citep{Lee1980,Loussaief2015,Kezirian1992}. Thus it stands as a unified approach which circumvents the necessity of two different analysis tools in different regimes, i.e. an image singularity based far field analysis \citep{Spagnolie2012,Lopez2014} and a singular perturbation analysis in the lubrication regime \citep{Ishikawa2006}.

The complete swimming problem is decomposed into a thrust problem (considering the case when the swimmer is held fixed and experiencing only a tangential surface velocity) and a drag problem (when it undergoes a rigid body motion with $\{\mathbf{V},\boldsymbol{ \Omega}\}$
and experiences hydrodynamic drag). In the $z$ direction, the force-free condition (\eqref{eq:force_free}) reduces to
\begin{subequations}
	\label{eq:force_torque_bal_zxy}
	\begin{equation}\label{eq:force_bal_z}
	F^\text{(Drag)}_{z}+F^\text{(Thrust)}_{z}=0, 
	\end{equation}
	\begin{equation}
	\qquad F^\text{(Drag)}_{x,T}+F^\text{(Drag)}_{x,R}+ F^\text{(Thrust)}_{x}=0 
	\end{equation}
	\begin{equation}
	\text{and}\quad	L^\text{(Drag)}_{y,T}+L^\text{(Drag)}_{y,R}+ L^\text{(Thrust)}_{y}=0.
	\end{equation}
\end{subequations}
Here, various hydrodynamic forces (torques) and velocity (angular velocity) components  are linearly related through various resistance coefficients (denoted with `$ f $') as: $F_z^\text{(Drag)}=f_z V_z,\;F^\text{(Drag)}_{x,T}=f_{x,T}\;V_x,\quad F^\text{(Drag)}_{x,R}=f_{x,R}\;\Omega_y, \quad L^\text{(Drag)}_{y,T}=f_{y,T}\;V_x\;$ and $\;L^\text{(Drag)}_{y,R}=f_{y,R}\;\Omega_y$.
The hydrodynamic resistance coefficients are functions of only the distance of the microswimmer from the wall $ (h) $ and the slip length $ (l_{\!_S})$, while the thrust force and torque are also dependent on the squirmer variables $ \beta$ and $\theta$. Once the solution of a particular `fundamental problem' (see \ref{sec:sol_detail} for detail) is found, the resistance coefficients, the thrust force and torque can be determined by the series summations in terms of the constants in the eigenfunction expansions \citep{Lee1980}.
Hence, the velocity components $ V_x, V_z $ and $ \Omega_y$ are easily obtained by solving equations~\ref{eq:force_torque_bal_zxy}  (a-c). 
\subsection{Reciprocal theorem for a microswimmer near a slippery surface}
\label{ssec:reciprocal}
The propulsive force and torque on the microswimmer can be alternatively determined  without solving the Stokes equation by utilizing the Reynolds Reciprocal Theorem (RRT) between two Stokes flows with the same geometry which has the general form \citep{Happel2012}
\begin{equation}\label{eq:reciprocal_gen}
\iint_{\partial S} \mathbf{n} \cdot \boldmath{\sigma '} \cdot \mathbf{u ''} = \iint_{\partial S} \mathbf{n} \cdot \boldmath{\sigma ''} \cdot \mathbf{u '},
\end{equation} 
where $ \mathbf{u'}, \boldsymbol{\sigma'} $ correspond to the swimming problem with a tangential squirming velocity and $ \mathbf{u''}, \boldsymbol{\sigma''} $ describe a complementary Stokes problem. Here $ \partial S $ is the boundary of the fluid domain.  

Previous works related to the motion of a microswimmer near a no-slip surface \citep{Crowdy2011,Crowdy2013,Mozaffari2016} have taken the advantage of vanishing fluid velocity at the plane wall to reduce the flow boundary in \eqref{eq:reciprocal_gen} to the swimmer surface only, i.e. $ \partial S=S_p.$  
However, it is not the case for a plane wall with a slipping boundary condition and we are left with  
\begin{equation}\label{eq:reciprocal_gen_area_split}
\iint_{S_p} \mathbf{n} \cdot \boldmath{\sigma '} \cdot \mathbf{u ''}+\underbrace{\iint_{S_w} \mathbf{n} \cdot \boldmath{\sigma '} \cdot \mathbf{u ''}}_\text{Wall slip contribution}  = \iint_{S_p} \mathbf{n} \cdot \boldmath{\sigma ''} \cdot \mathbf{u '}+\underbrace{\iint_{S_w} \mathbf{n} \cdot \boldmath{\sigma''} \cdot \mathbf{u'}}_\text{Wall slip contribution},
\end{equation} 
where $ S_w $ is the surface of the slippery plane wall.
It was shown  \citep{Loussaief2015} that the contributions of the extra terms due to non-vanishing fluid velocity at the plane wall, appearing in both sides of the \eqref{eq:reciprocal_gen_area_split}, become equal,  i.e.
\begin{equation}\label{eq:reciprocal_gen_area_slip}
\iint_{S_w} \mathbf{n} \cdot \boldmath{\sigma '} \cdot \mathbf{u ''}  = \iint_{S_w} \mathbf{n} \cdot \boldmath{\sigma''} \cdot \mathbf{u'}.
\end{equation} 
and we subsequently obtain 
\begin{equation}\label{eq:reciprocal_gen_area_particle}
\iint_{S_p} \mathbf{n} \cdot \boldmath{\sigma '} \cdot \mathbf{u ''} = \iint_{S_p} \mathbf{n} \cdot \boldmath{\sigma''} \cdot \mathbf{u'}.
\end{equation} 
Next we choose the complementary problem as the motion of a spherical particle near a slippery plane wall where the particle has a translational velocity $ \mathbf{U''} $ and rotational velocity $ \boldsymbol{\Omega''}$. Utilizing the force and torque-free conditions of the microswimmer and employing the boundary condition (\eqref{eq:BC_swimmer}) on the swimmer surface, finally \eqref{eq:reciprocal_gen} is simplified to
\begin{equation}\label{eq:reciprocal_swimmer}
\mathbf{F''}\cdot \mathbf{U}+\mathbf{T''}\cdot \boldsymbol{\Omega}=-\iint_{S_p} \mathbf{n} \cdot \boldsymbol{ \sigma''} \cdot \mathbf{u^s}\,dS,
\end{equation}
where $\mathbf{F''} $ and $ \mathbf{T''} $ represent the force and torque on the spherical particle in the complementary Stokes problem. The above equation suggests that the translational and rotational velocities of the microswimmer can be found by knowing the surface tangential velocity and the solution of the complementary Stokes problem, thereby bypassing the detailed solution of the swimmer problem.

\section{Results and discussions}
 Towards investigating the microswimmer trajectories we solve the following dynamic system
\begin{subequations}\label{eq:2}
	\begin{gather}
	\frac{dx(t)}{dt}=V_x,   \qquad   	\frac{dh(t)}{dt}=V_z,   \qquad   	\frac{d\theta(t)}{dt}=\Omega_y.    \tag{\theequation a-c}
	\end{gather}
\end{subequations}
Here we do not consider any stochastic motion due to translational or rotational diffusion and the trajectories are computed based on deterministic forces only. 
Close approach of a microswimmer towards the wall often leads to swimmer crashing against the wall and the subsequent  motion becomes untraceable. To overcome this problem, we employ an additional short range repulsive force of the form \citep{Spagnolie2012} $: \mathbf{F}_\mathrm{rep}= \dfrac{\alpha_1 \exp{(-\alpha_2 \, \delta)}}{1-\exp{(-\alpha_2 \, \delta)}} \mathbf{e_z} $. Following \citep{Spagnolie2012} the parameter values $ \alpha_1=100,\alpha_2=100$ are chosen to prevent the swimmer coming not closer than a distance of $\sim 0.01$ times the swimmer radius from the wall. Such forces originate from the nanoscale interaction between the swimmer body and the wall surface, especially in physiological conditions \citep{Klein2003}. Diverse forms of repulsive forces have been employed in the literature \citep{Li2014,Walker2019,Hu2015,Katuri2018}. 
It is to be noted that the use of short-range repulsive force at the wall destroys the puller-pusher duality during time reversal \citep{Ishimoto2017,Walker2019}.

First we illustrate the wall-slip mediated alterations in the translational and rotational velocity components of different types of squirmers having a spherical cell body.  Subsequently, the resulting trajectories are elucidated and different phase transitions of swimming states are discussed. Considering typical microswimmer radius in the range of $ 1-100\; \mathrm{\mu m} $ and in view of the experimentally observed dimensional slip lengths \citep{Huang2008,Zhu2001,Tretheway2002}, we take the dimensionless slip length $(l_S)$ in the range of 0 to 10.

\subsection{Swimming velocity alterations}
\label{ssec:vel_result}

\begin{figure}[!htb]
	\centering
	\includegraphics[width=1\textwidth]{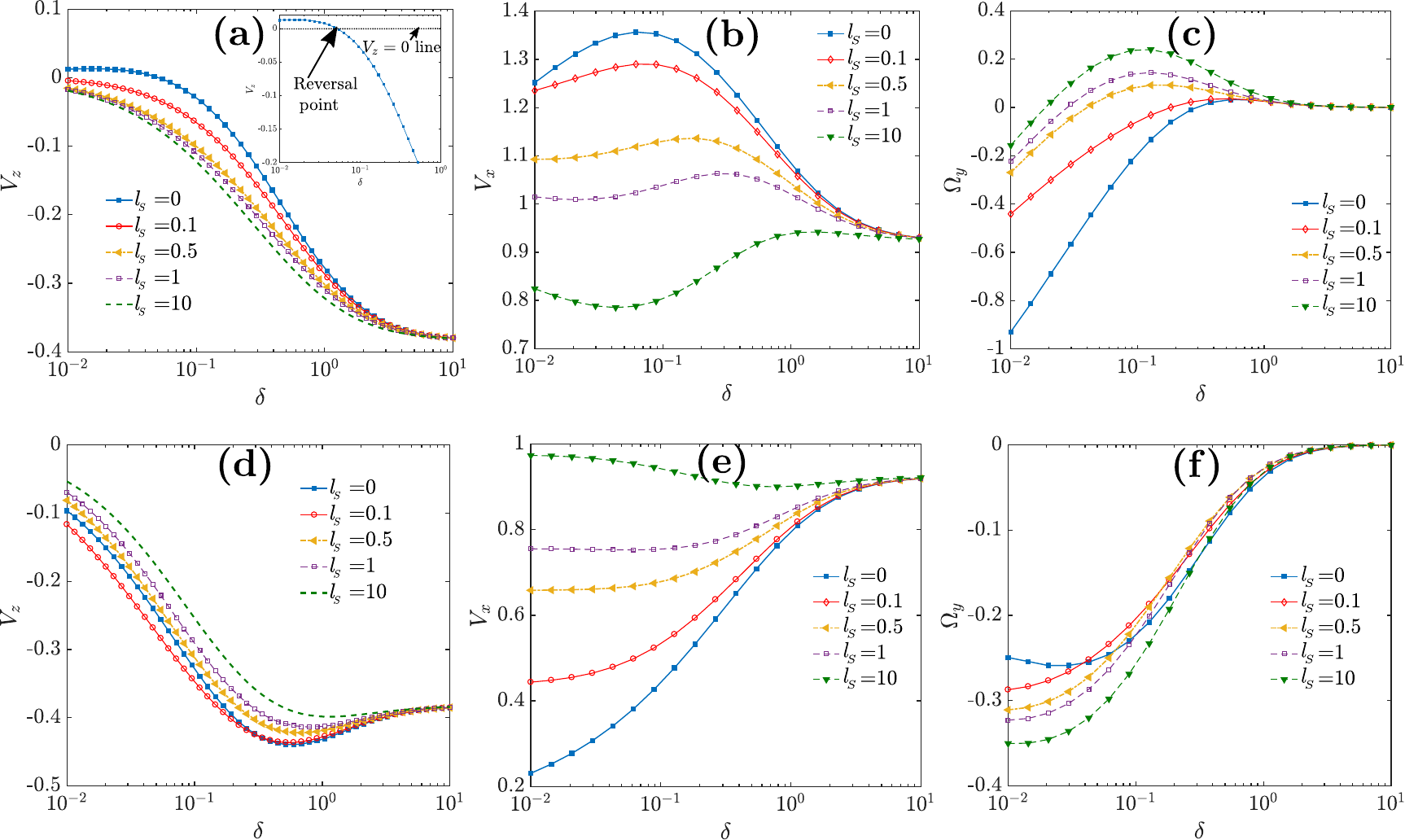}
	\vspace{-2ex}
	\caption{Velocity components of a microswimmer vs. smallest separation distance from the wall $ (\delta=h-1) $ for different slip length $ (l_{\!_S}) $. Subfigures (a), (b), (c) are for a puller and (d), (e), (f) are for a pusher, all having a squirmer parameter value, $ |\beta|=4$ and angular orientation, $ \theta=\pi/8.$ In the inset of (a), a magnified view of the no-slip case is shown.} 
	\label{fig:vel}
\end{figure}
Figure~\ref{fig:vel} portrays the effects of slip length $ (l_{\!_S}) $ on the velocity components $ (V_z,V_x,\Omega_y) $ at different separation distances of the microswimmer from the wall $ (\delta) $. In many situations the said effects turn out to be different in nature, if not opposite, for the puller and pusher type microswimmers. To understand the physical origin of the same, we first look into the fundamental differences in the propulsion mechanisms and surrounding flow patterns associated with these two types of swimmers in an unbounded domain.  Pullers have their flagella ahead of their cell body and the thrust generated by the flagella is cancelled by the cell body at the back. Thus, at large length scales, a moving puller squirmer gives rise to a contractile dipolar flow in the surroundings (figure~\ref{fig:schematic}(b-i)). In contrast, pushers have their flagella at the back and the thrust is generated from behind, thereby mimicking a extensile dipolar flow at large length scales (figure~\ref{fig:schematic}(b-ii)). Also beyond a sufficient stirring action created due to the second squirming mode $(|\beta|>1)$, a pair of circulation rolls, symmetric about the direction of motion, is observed from a co-moving frame, but at different locations, behind the cell body for a puller but ahead of the cell body for a pusher \citep{Magar2003,DeCorato2015}.  If we now consider a case when swimmers of both type are pointing towards the wall, the effect of wall reflected flow is first sensed by the flagella in case of a puller and the thrust created by their breast-stroke action gets affected due to a confinement. In contrast, the frontal cell body of the pusher will be the first to experience the distorted flow, which must be adjusted by a modified pushing action by their posterior flagella. Evidently, the stress distribution around the swimmer will also face modifications based on the relative position of the cell body and the flagella and the corresponding flow adjustments created by them. The presence of wall slip further complicates the scenario by mediating the propulsive action of the swimmer and simultaneously modifying the hydrodynamic resistance of a finite size cell body moving near the wall.

In figure~\ref{fig:vel}(a), the effect of wall slip on the wall normal velocity component $ V_z $ is shown for a puller microswimmer with locations ranging from very close to far away separations from the plane wall. During the wall normal movement of a spherical particle, it is known that the hydrodynamic drag force shows a rapid rise, almost varying inversely with the distance $\delta$ \citep{Cooley1969}. Employing lubrication theory analysis for $\delta \ll 1$, \citep{Hocking1973} showed that the presence of wall slip results in a logarithmic increase in $f_z$ with $\delta$ which can be quantified as:
$f_z \sim \dfrac{1}{3 l_{\!_S}}\left[ \left(1+\dfrac{\delta}{6 l_S}\right)\ln \left(1+\dfrac{6 l_S}{\delta}\right)-1 \right]$, while it becomes inversely proportional to the slip length $ (l_{\!_S}) $. Further, in situations of high slip length relative to the wall separation distance $(l_S \gg \delta)$, we get 
$f_z \sim  \left(\dfrac{\delta}{3 l_S}\right)\ln \left(\dfrac{6 l_S}{\delta}\right)$. While considering the behaviour of microswimmer velocity $ V_z $, we also have to examine the thrust force variations shown in figure~\ref{fig:prop_force}. Figure~\ref{fig:prop_force}(a) demonstrates that  a puller type microswimmer experiences a decreasing thrust force when the wall is slip-free and it becomes negative after the separation exceeds a certain value. This is responsible for velocity reversal in figure~\ref{fig:vel}(a). With increasing slip length this reversal in $ V_z $ ceases to occur and it indicates a wall-approaching trend of the microswimmer for all separation distances, being consistent with the thrust force variations. Also, in the cases of rising magnitude of $|V_z|$ with increasing slip, the decrease in hydrodynamic drag becomes a dominant factor. However, the resultant impact of $ \delta$ and $l_S$ on the magnitude of $ V_z $ is determined by the competitive effects of those parameters on both the drag $ (F_z^\text{(Drag)}) $ and thrust $ (F_z^\text{(Thrust)}) $ forces.

Figure~\ref{fig:vel}(d) depicts that for a puher-type swimmer, the  variations in $ V_z $ are non-monotonic in both $ \delta $ and $ l_{\!_S}$, although the thrust force is monotonic (figure~\ref{fig:prop_force}(d)). Observing figure~\ref{fig:pusher_vz_suplle}, we find that at very low wall separations $ (0.01 \lesssim \delta \lesssim 0.1) $, there exists an intermediate slip length, $l_S\sim 0.1 $ for which the wall bound velocity becomes maximum in magnitude. This again indicates a considerable importance of hydrodynamic drag force variations on the swimmer velocity. Notably, the above observations for a pusher are in sharp contrast to the characteristics of a force dipole pusher, which, in the far field analysis, always tends to get attracted to the wall with increasing slip length \citep{Lopez2014}. Such disagreement arises from the fact that in a force dipole model, the contributions from the higher order singularities arising from a finite size cell body or the fore-aft asymmetry of the swimmer, are not taken into account \citep{Spagnolie2012}. In addition to these far-field effects, in the present case, the near field hydrodynamic interaction also plays its role in modifying the swimmer velocity. 

Figure~\ref{fig:vel}(b) shows that the effects of wall slip on the wall-parallel velocity component $ (V_x)$ of  a puller are much pronounced for low separation distances. As the swimmer moves away from the wall, there exists a specific wall gap where the $ V_x $ becomes maximum for that slip length. With increasing slip length, the maximum velocity occurs at greater distances from the wall.   
Moreover, the slip-induced suppression $ |V_x-V_{x,\text{No slip}}| $ is maximum for an intermediate wall-swimmer distance of nearly $ \delta \approx 0.1$ and it gradually becomes vanishingly small as the microswimmer reaches the velocity of a solitary swimmer at distances nearly 10 times the swimmer radius. It is also interesting to observe that the wall-parallel velocity remains higher than the unbounded medium velocity, i.e. $ (V_x \to \cos(\theta)) $ for all wall separations, until the slip length becomes high enough (e.g. $ l_S=10$) to cause $ V_x $ falling below the unbounded medium velocity. Fall in swimming velocity with increasing $ l_S $ is in contrast to the intuition since a spherical particle translating parallel to the wall  experiences lower hydrodynamic resistance in the presence of slip \citep{Davis1994}, which for small slip lengths $ (l_S \ll 1) $ is quantified by the following expression \citep{Loussaief2015}:
$f_{x,T}/(-6\pi) \simeq -\dfrac{8}{15}\log \left(\dfrac{\delta+l_S}{1+\delta+\l_S}\right)+\dfrac{0.9543+\delta+l_S}{1+\delta+l_S}$. 
 To address this apparent anomaly, we consider the fact that the microswimmer velocity is dependent on a coupled effect of the hydrodynamic resistances to the simultaneous translational and rotational movement parallel to the wall as well as on the thrust generated due to the propulsive action. Both the propulsive thrust force $ (F_x^\text{(Thrust)}) $ and torque $(L_y^\text{(Thrust)})$ are affected by wall slip, as shown in figures~\ref{fig:prop_force} (b) and (c), respectively. It is observed that the increase in $F_x^\text{(Thrust)}$ with reducing swimmer-wall distances in the no-slip case is now slowed down by the wall slip effect. Moreover, when the slip length is much higher, the thrust force even gets reduced with the swimmer approaching more towards the wall. We also observe a competitive nature of the effects of wall separation and the slip length on the thrust force experienced by the swimmer when it comes close to the wall.
In addition, the wall slip acts to reduce the counter clockwise (CCW) propulsive torque and even makes it clockwise (CW) for extremely high slip lengths, but only for an intermediate range of wall separations (see \ref{fig:prop_force}(c)). A positive (CW) propulsive torque also contributes in reducing $ V_x$ which can be confirmed from the final expression of $ V_x $ in equation~\ref{eq:Vx_Wy_coupled}(a). 

The sign of the angular velocity of a microswimmer near a wall $ (\Omega_y)$ is determined by two main opposing physical mechanisms arising from the propulsive action: the torque whose direction depends on the direction of the surface flow at the surface of the swimmer and the torque arising from the wall-parallel linear motion with $ V_x$ whose direction remains clockwise due to forward movement.
Again, the final expression of $ \Omega_y$ in equation~\ref{eq:Vx_Wy_coupled}(b) suggests that the  hydrodynamic resistive factors are intrinsically coupled with the thrust force and torque and take part in deciding the resultant  rotation direction. The effect of wall slip on the various terms contributing to the rotation rate are shown in figure~\ref{fig:Torque_3_terms}.  This shows that increasing wall slip drastically alters $ T_1 $ and $ T_3$, while changes in $ T_2 $ are not significant. In effect, for a puller swimmer, the wall-slip exerts a strong opposing torque to the CCW rotating swimmer, which for high slip lengths causes it to rotate towards (CW) the wall, before reaching the bulk zero rotation state at a distance of nearly one radius away from the wall.
\begin{subequations}\label{eq:Vx_Wy_coupled}
		\begin{gather}
		V_x=\dfrac{F_x^\text{(Thrust)} f_{y,R}- L_y^\text{(Thrust)} f_{x,R} }{f_{x,R}f_{y,T}-f_{x,T}f_{y,R}}   \quad \text{and} \quad   	\Omega_y=\dfrac{\overbrace{L_y^\text{(Thrust)} f_{x,T}}^{T_1}- \overbrace{F_x^\text{(Thrust)} f_{y,T}}^{T_2} }{\underbrace{f_{x,R}f_{y,T}-f_{x,T}f_{y,R}}_{T_3}}.   	\tag{\theequation a-b}
		\end{gather}
\end{subequations}

Variation of the thrust force $(F_x^\text{(Thrust)})$ for a pusher (see figure~\ref{fig:prop_force}(e)) is highly complex and non-monotonic in nature. However, the corresponding wall-parallel velocity component $(V_x)$ shows a trend of getting escalated with slip length (see figure~\ref{fig:vel}(e)), a phenomenon which is exactly opposite to that of a puller. The CCW thrust torque on the microswimmer gets reduced by the wall slip for all separations (see figure \ref{fig:prop_force}(f)). Again this fixed trend is not followed by the rotation rate $\Omega_y$. 
Figure \ref{fig:vel}(f) demonstrates that for a pusher swimmer located very close to wall $ (\delta \lesssim 0.04)$, the wall slip forces it to rotate away from the wall (CCW). However, beyond this distance the slip-induced CCW rotation either gets escalated or suppressed, depending on a subtle interplay between slip and wall separation. This is again in contradiction to the previously studied far-field behaviour of a force dipole swimmer \citep{Lopez2014}, due to the similar reasons discussed for $ V_z.$

\subsection{Modulations in near-wall swimming trajectories}
\label{ssec:trajectory}
\subsubsection{Neutral squirmer}
\label{sssec:netral_traj}

\begin{figure}[!htb]
	\centering
	\includegraphics[width=0.75\textwidth]{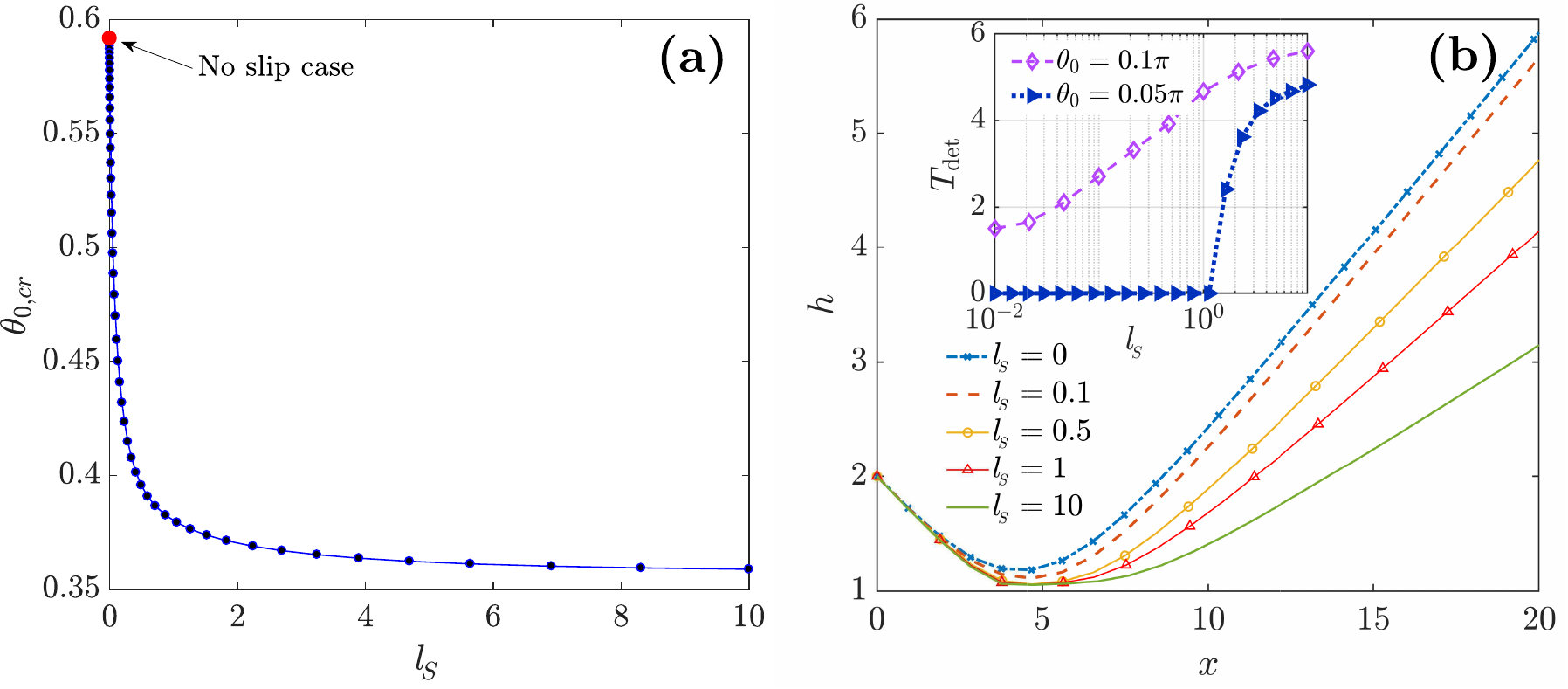}
	\vspace{-2ex}
	\caption{ (a) Critical initial orientation $ (\theta_{0,cr}) $ vs. slip length $ (l_{\!_S})$. (b) Trajectory with initial orientation, $ \theta_0=0.1\pi$. In the inset the variation of dimensionless detention time with slip length is shown. In both the subfigures the initial height, $ h_0=2$ is taken.} 
	\label{fig:Netrual_trajectory}
\end{figure}

	A neutral squirmer which is characterized by $ \beta=0 $ and gives rise to a quadrupolar flow field around the swimmer in an unbounded domain and thus vorticity is absent in the velocity field \citep{Zhu2012,Zhu2011}. In the absence of the wall slip, for an initial orientation  away from the wall or even with a small tilt towards the wall, the a neutral swimmer does not show any tendency to move towards the wall and escapes from the wall  along a straight line with the final orientation angle reaching an asymptotic constant value $(\theta_f)$. The scenario changes for moderate values of the initial orientation angles towards the wall. As it moves towards the wall, its director gradually points away from the wall due to a net CCW torque arising from near-field hydrodynamic effect which subsequently forces the swimmer to attain a negative orientation angle $\theta_0<0$. In effect, the normal velocity component becomes positive wall-normal velocity $ V_z>0 $ leading to escape of the swimmer away from wall. Beyond a critical initial orientation, $ \theta_{0,cr}$ the swimmer eventually collides with the wall, remains in the wall-adjacent region for some time and finally escapes with a final orientation equal to its initial one, $ \theta_f\approx\theta_0$. When the swimmer has an initial height of $ h_0=2 $, this critical angle for descending to a height below $ h=1.05$ had been reported to be $ \sim 0.4 $ by \cite{Spagnolie2012}. 
How the wall slip intervenes in the dynamics of a Neutral swimmer? Answering this, we show in figure~\ref{fig:Netrual_trajectory}(a) that the critical condition for the transition of a scattering trajectory to a colliding one, is highly influenced by the wall slip as manifested through the decrease in this critical initial tilt angle  with increasing wall slip length. The critical angle, with a cut-off distance for collision as, $ h=1.01$, shows a drastic decrease from the no slip case $ (l_{\!_S}=0, \theta_{0,cr}=0.59) $ to $ \theta_{0,cr}=0.38 $ for $ l_{\!_S}=1 $ and finally reaches an asymptote of $ \theta_{0,cr} \to 0.36 $ for higher values of the wall slip length.

Wall-bound detention time of microswimmers of different type has been observed previously near no-slip surfaces \citep{Li2014,Schaar2015}.  We quantify the detention time $ (T_\text{det}) $ by the time interval during which the microswimmer remains below a distance of one-tenth of its diameter. For a typical microswimmer diameter of $10$ to $30$ $ \mu m $, this cut-off distance remains consistent with the experimental evidence of \textit{E.Coli} cells remaining at a distance of $<1-3 $ $\mu m$ near a wall \citep{Drescher2011} for an extended time. Figure~\ref{fig:Netrual_trajectory}(b) shows that although the escaping nature of the microswimmer motion is preserved even with a very high slip length,   the detention time for a neutral swimmer gets increased significantly with the slip length $ (l_{\!_S}) $ (see the inset). Interestingly, for small tilt angles towards the wall (e.g. $\theta_0=0.05\pi$), where the detention time is negligible near a no-slip wall, the increasing slip length beyond a critical value is found to impart a high detention time.

\subsubsection{Puller squirmer}
\label{ssec:puller}

While swimming near a no-slip wall, the director $ (\mathbf{e}) $ does not face a hydrodynamic rotation relative to the wall if the strength of the vorticity generation term $ (\propto \beta) $ in the squirmer surface velocity is not sufficiently high \citep{Ishimoto2013}. As $ \beta $ crosses a critical value, the hydrodynamic torque imparts an extra rotation of the puller towards the wall and stable swimming state parallel to the boundary takes place favoured by an initial tilt towards the wall, i.e. $ \theta_0 >0$, while the wall effects are negligible for $ \theta_0 <0 $ \citep{Li2014}.  
We found that, even in the presence of wall slip, the wall bound attraction of swimmer is non-existent for small values of the squirmer parameter $ \beta \lesssim 2.75$ with any initial director orientation.  Similar to the neutral squirmers, we observe that the escaping nature is affected in the sense that the minimum height reached by the swimmer gets reduced and the detention time near the wall is increased with rising slip lengths. 
  
\begin{figure}[!htb]
	\centering
	\includegraphics[width=1\textwidth]{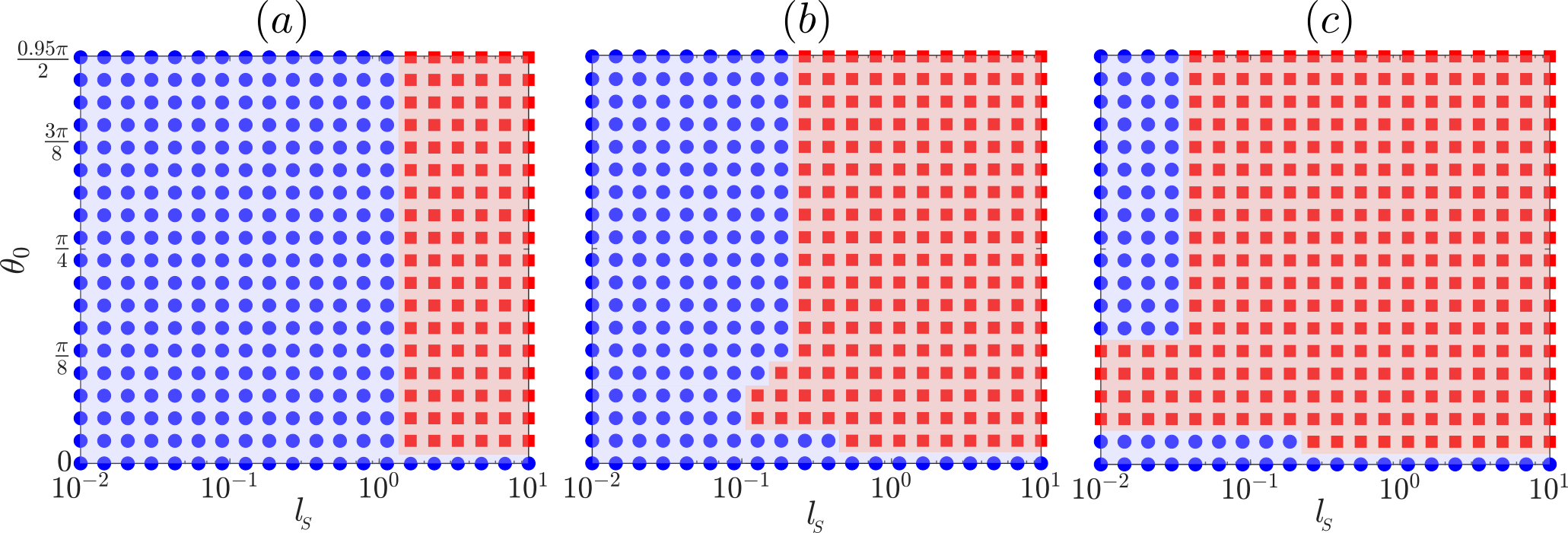}
	\vspace{-2ex}
\caption{Phase maps of the final swimming states of puller microswimmers in the $ (l_{\!_S},\theta_0) $ plane. Subfigures (a),(b) and (c) correspond to squirmer parameters $ \beta=3, 4.5 $ and 5, respectively.   The `blue' circles and `red' squires correspond to the escaping and trapping states, respectively. In all the presented case, the $ l_{\!_S} = 0.01 $ case gives swimming state similar to a no slip wall. The illustrations are with an initial launching height of $h_0=2$.    } 
\label{fig:Phase_trajectory_repulsion}
\end{figure}

  The scenario changes as $ \beta $ crosses this limiting value and the wall slip triggers a transition of swimming states from wall-escape to wall-entrapment, as summarized in the phase diagrams of ~\ref{fig:Phase_trajectory_repulsion}(a)-(c). 
 {Phase diagrams depicting diverse trajectory characteristics of squirmers \citep{Uspal2015} as well as three-sphere microswimmers \citep{Daddi-Moussa-Ider2018} near a no slip surface have been constructed previously on the basis of the release height and orientation. In stark contrast, in the present study we represent unique phase diagrams elucidating the immense contribution of hydrodynamic slip length in deciding the resulting trajectory.}
    All the escaping trajectories have been confirmed from long time simulations with a cut-off   distance of $h_\mathrm{escape}=15$ \citep{Ishimoto2013}. The trajectories starting from a launch angle $\theta_0 \sim \pi/2$, which cause  direct impact \citep{Spagnolie2012} and become computationally demanding, have been excluded from computation by limiting the initial orientation angle in the range $ \theta_0 \le 0.95 \times \pi/2$.
    Presence of wall slip severely modifies the near field hydrodynamic interaction, as observed during the discussions of velocity components in section~\ref{ssec:vel_result}. As a consequence, low values of the initial tilt angle which do not lead to any trapping state near a no-slip wall, are found to be sufficient for the swimmer to attain that extra rotation towards the wall which favours a state transition from escaping to wall-bound trapping, either in the form of periodic oscillations or steady state sliding.
   We would like to draw the attention of the reader that in no-slip case itself, previous studies have  reported wall-entrapment where the wall is equipped with a repulsive force of different forms \citep{Lintuvuori2016,Ishimoto2017}. As a confirmation that the swimming state transitions are exclusively caused by hydrodynamic slippage, we observed that many trajectories which showed escaping nature without $\mathbf{F}_\text{rep}$ are not characteristically affected due to the presence of the short range repulsive force alone, and interfacial slip beyond certain limit $(l_{\!_S,\text{cr}})$ is essential to cause the required reorientation for swimming state transition.

\begin{figure}[!htb]
	\centering
	\includegraphics[width=1.1\textwidth]{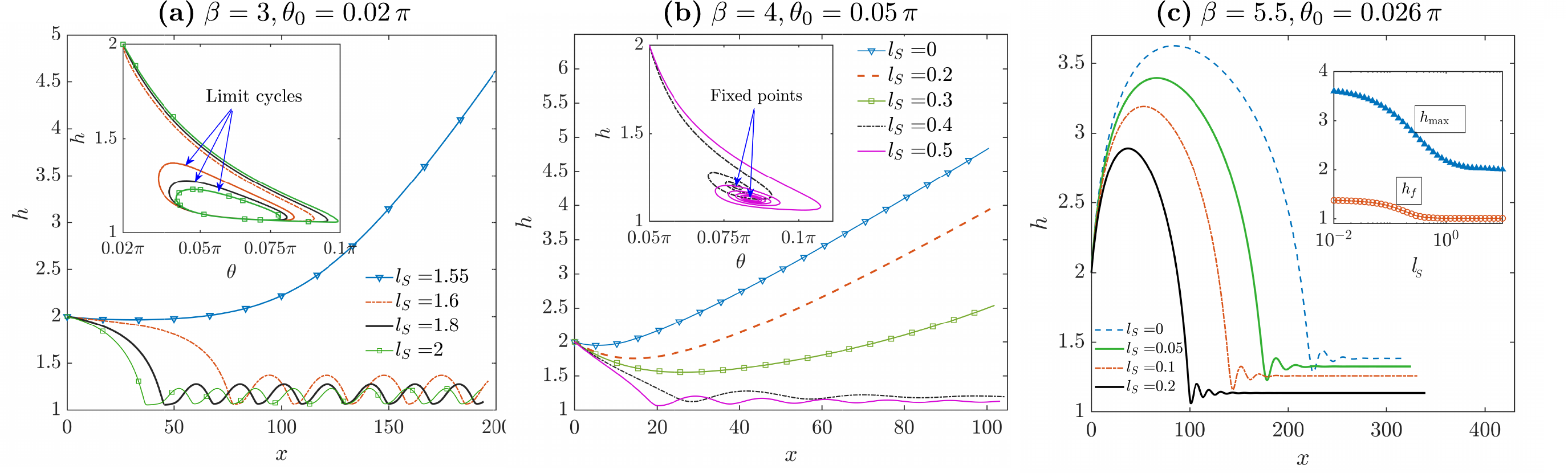}
	\vspace{-4ex}
 	\caption{Different characteristics of the slip-induced swimming state transitions for different squirming parameter for puller swimmers $(\beta)$. The corresponding parameters are shown against each subplot. The insets to subplots (a) and (b) correspond to the trajectory in the phase space, while the inset to subplot (c) depicts the variations maximum attained height $ (h_\text{max} )$ and final steady state sliding height $ (h_f) $ with slip length $(l_{\!_S}) $.} 
\label{fig:puller_beta_3_4}
\end{figure}

      For moderate values of $ \beta $ (i.e. $ 2.75 \lesssim \beta \lesssim 3.75 $), as the initial swimmer orientation departs slightly from the wall parallel direction $ (\theta_0=0)$, the transition phenomenon occurs sharply at a fixed value of the slip length $ (l_{{\!_S},\text{cr}}) $, without showing any non-monotonic dependence on $ \theta_0$. In this case, after the swimmer collides with the wall, it gets a CCW rotation due to a net near-field hydrodynamic repulsive torque and shows a tendency to escape. However, due to an attractive effect of the near-wall forces, its normal velocity changes its sign $ (V_z<0) $ after reaching a certain height and again swims towards the wall. Finally, a periodic oscillatory trajectory results (see figure~\ref{fig:puller_beta_3_4}(a) for a representative example) in and limit cycles emerge in the phase plane of the dynamic system. We also observe that the amplitudes of the present periodic oscillations get decreased with rising slip lengths.

With further increase in $\beta$, i.e. $ \beta \gtrsim 3.75 $, the trapping states become damped oscillatory in nature beyond a critical slip length, as portrayed in figure~\ref{fig:puller_beta_3_4}(b),(c)  and eventually steady state stable swimming takes place with a fixed height and orientation $ (h_f,\theta_f)$. These final swimming states gives rise to fixed points in the phase map of the dynamic system of $ (\dot{h},\dot{\theta})$. 
The transition of periodic to damped amplitude trapping with increase in $\beta$ suggests that during the combined influence of the  torques due to hydrodynamic slip and that due to the circulating flow pattern emerging from the propulsive action,  if the former one is more dominant than the other, damping effect on swimmer oscillations is less prominent.  This condition further arises in specific cases of pusher swimmer characteristics, to be discussed subsequently.

As observed in figure~\ref{fig:puller_beta_3_4}(c), for a high value of $ \beta $ (e.g. $ \beta=5.5$) the swimmer initially moves away from the wall much above the initial height but gets attracted towards the wall due to high reorientation torque imparted by pronounced contribution of the stresslet term in swimmer surface velocity and shows some small amplitude oscillations in height before getting trapped.
 Additionally, the maximum height reached during the first bouncing motion $(h_\text{max})$ and the final sliding height $ h_f $ both get reduced due to wall-slip. Also the longitudinal distance travelled before coming too close to the wall gets decreased.
 This effectively portrays the wall slip effect as a strong influencing parameter to cause trajectory transition even if the swimmer reaches a height much above the wall. On the other hand, the instance of decreasing in minimum swimmer-surface distance owes its origin in the nature by which the balance between the near-wall hydrodynamic attraction and short-range repulsion is interfered by the hydrodynamic slippage.

  Interestingly, with increase in $ \beta$ (see  figures~\ref{fig:Phase_trajectory_repulsion}(b),(c)) for a band of low $\theta_0$ values, the swimming state transition occurs even with zero wall slip. However, beyond a critical high value of the slip length, only the stable trapping states exist for all initial tilt angles considered and the non-monotonicity with $ \theta_0 $ vanishes. This renders the critical slip length dependent on initial launching angle, i.e. $ l_{\!_S}= l_{\!_S}(\theta_0)$. Some illustrative trajectories are shown in figure~\ref{fig:T_R_beta_5_Lambda_0p029764_non_monotonic_th0}(a). It depicts that for some initial orientations in the intermediate range $ 0.05 \pi \lesssim \theta_0 \lesssim 0.16 \pi $, the swimmer travels a maximum vertical distance almost as high as the initial height, $h_0=2$, but thereafter sensing the slip in the wall, it follows a damped amplitude oscillatory motion and finally slides parallel to the wall at a fixed height, $h_f=1.464$ and positive angle (see figure~\ref{fig:T_R_beta_5_Lambda_0p029764_non_monotonic_th0}(b)), $\theta_f=0.266.$ Beyond this $ \theta_0$, the swimmer bounces on the wall,  thereafter travels to height $ h>h_0 $ which is beyond the reach of the reorientation torque of the slippery wall and finally escapes away from the wall.

  Before a full transition of escape to trapping occurs, rising wall slip affects the trajectories by increasing the wall bound detention time $(T_\text{det})$ for both puller and pushers, similar to the previously discussed case of a neutral one. A representative behaviour with $\beta=3$ is described in  	figure~\ref{fig:Detention_time_beta_3_vary_ls_th0_0p1pi_T_R}. In these cases, even after the collision with the wall, the torque on the swimmer is not sufficient to impart a sliding type motion near the wall. Rather, the swimmer escapes the wall-adjacent region after gliding along the wall for a finite detention time.  
  Thus, the wall-bound detention time of the neutral and puller microswimmers can be controlled by suitably tuning the wall slip length, thus facilitating the formation of bio-aggregates such as biofilms near a wall \citep{Watnick2000}. 

\subsubsection{Pusher squirmer}
\label{sssec:Pusher}
The swimming states of a pusher swimmer also show wall-bound trapping nature beyond a critical slip length $(l_{s,\text{cr}})$. We have previously observed that for pullers the emergence of slip-instigated trapping comes into existence only with an initial tilt slightly towards the wall $ (\theta_0>0) $ while negative initial angles $ (\theta_0<0) $ give rise to escaping states only.  However, in case of pushers, we find that these stable trapped states exist even for small initial tilt away from the wall, but a further increase of tilt away from the wall results in escaping states only (see figure~\ref{fig:Trajectory_beta_minus_5_TH0_minus_0p165_0p_248} for an example).

 \begin{figure}[!htb]
	\centering
	\includegraphics[width=0.55\textwidth]{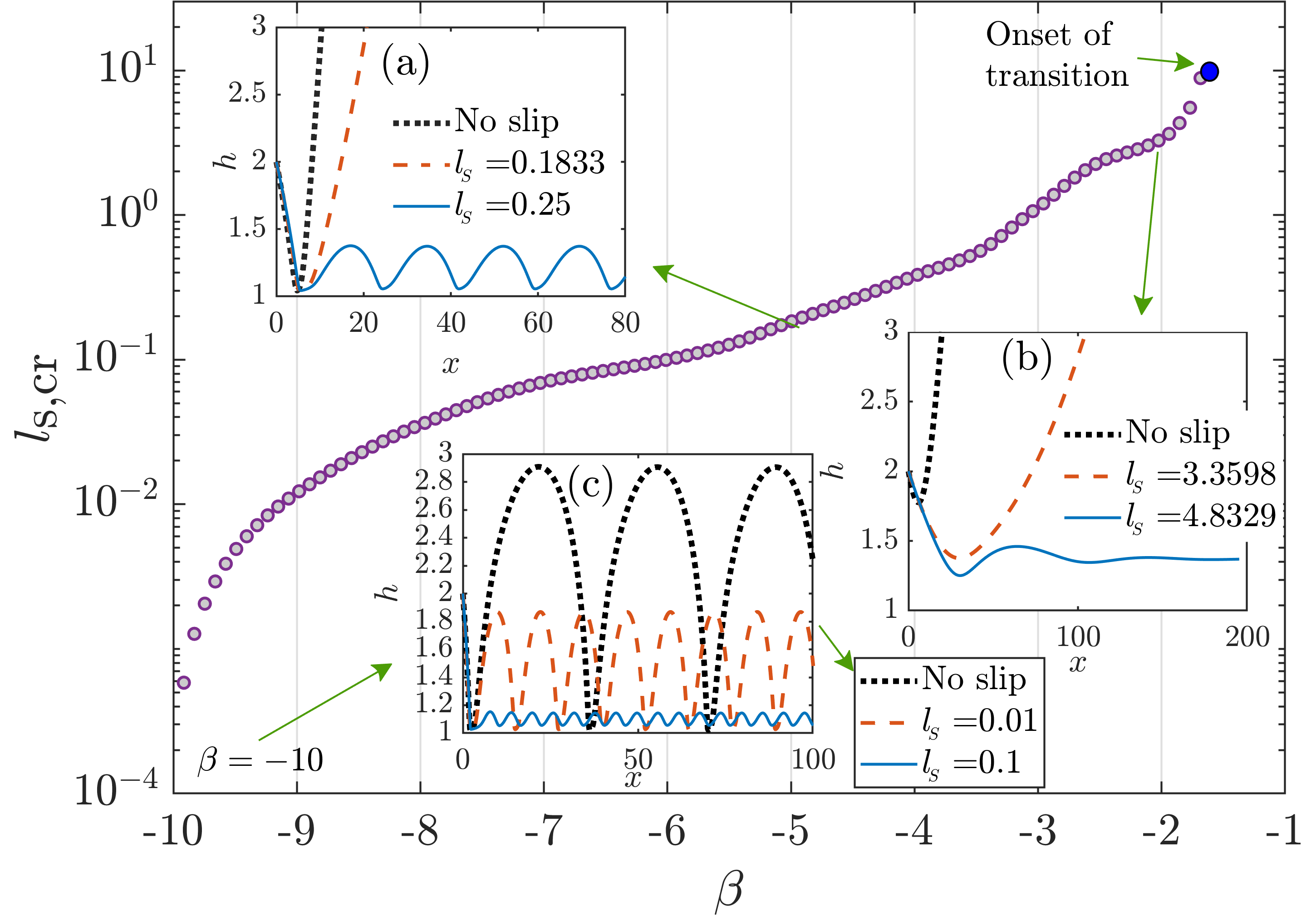}
	\vspace{1.5ex}
	\caption{Critical slip-length $(l_{s,cr})$ vs. squirmer parameter $(\beta)$ for pusher type swimmers with initial conditions $ (h_0,\theta_0)=(2,0) $. In the insets (a),(b) the transition behaviours are highlighted for  $ \beta=$ -5 and -2, respectively, while the inset (c) describes the characteristic changes in the periodic oscillations with increasing slip length for $ \beta=-10$. The onset of transition takes place for $ \beta \sim -1.6 $ and is denoted by a `blue' marker. } 
	\label{fig:Pusher_beta_vs_Ls_cr}
\end{figure}
In contrast to the pullers, the non-monotonicity of this critical slip length with initial angular orientations (as shown in figure~\ref{fig:Phase_trajectory_repulsion}), has not been observed in the case of pushers. Beyond this critical slip length, the swimmer slides along the wall for any initial orientations in the range of $\theta_0$ as discussed above. Figure~\ref{fig:Pusher_beta_vs_Ls_cr} depicts the consolidated effect of the wall slip on the pusher swimmers for a range of squirmer parameter $-10 \le \beta \le -2.$ The onset of transition from an escaping to damped amplitude oscillation takes place for $ \beta \sim 1.6 $ with a high value of slip length $( l_{\!_S}).$  The oscillations become periodic in nature for $|\beta| \gtrsim 5.$   This is again opposite to the trend of puller microswimmers which exhibit periodic and damped amplitude oscillations for low and high values of the same parameter, respectively. This difference can be related to the contrast in the near field hydrodynamic effects brought in by the fluid slip at the wall, as discussed in section~\ref{ssec:vel_result}. Realization of slip-mediated trapping for pushers with low $ |\beta| $ is of utmost importance in view of their non-existence near a no-slip wall.
 
For a significantly high strength of $|\beta|$, the periodic oscillations are present even in the absence of wall slip, as illustrated for $\beta=-10$ in the inset (c) of figure~\ref{fig:Pusher_beta_vs_Ls_cr}.   Here also, the slip has an important role to play in the form of increasing the frequency of oscillatory height and reducing their amplitudes. This, in turn, allows the swimmer to slide along the wall, maintaining a minimal wall separation as dictated by the balance of the wall repulsive force and the hydrodynamic force for a high slip length.

Starting from a same initial orientation, during the final states of a damped cyclic swimming, the puller swimmers point towards the wall (quantified by a positive angle) while pushers show the opposite trend (please refer to figure~\ref{fig:Angle_comparison}). This difference in sign (but not exactly opposite in magnitude) can be attributed to the differences in the physical mechanisms providing the reorientation torque during the trapping states and deciding the actual orientation angle at a particular height, which was also shown to give rise different rotation rates for the two type of swimmers, as described in figures~\ref{fig:vel}(c) and (f).  

 The slip modulated wall-bound  motion of squirmers shares a characteristic feature with the recent experimental investigation of \citep{Ketzetzi2018} who found a prominent tendency of the diffusiophoretic Janus colloids to self-propel adjacent to hydrophobic surfaces having high slip lengths and to occasionally leave the surface followed by a steady swimming in wall-proximity.
 In view of the similarity of the surrounding flow patterns in an unbounded domain, the behaviour of such chemically active synthetic particles has been previously mapped onto the squirmer model \citep{Michelin2014}.  However, near a confinement, the distribution of the chemical species around a self-diffusiophoretic particle gets modified which in turn affects the slip velocity at the particle surface \citep{Mozaffari2016,Uspal2015a}, in contrast to the squirmers which have a prescribed surface velocity. Thus a qualitative matching of the respective swimming states calls for an explicit account of the dependence of hydrodynamics interaction and chemical species distribution \citep{Popescu2018}.
 
 \section{Conclusions}
 \label{sec:conclusion}
 
To summarize, we have adopted a `squirmer' model to mathematically describe the swimming characteristics of microorganisms near a plane wall with hydrodynamic slippage. In the low Reynolds number regime, the governing fluid flow equations are solved by employing an exact solution technique in bispherical coordinate system and the hydrodynamic slippage at the wall has been modeled using the Navier slip boundary condition. This provided us a unified platform to investigate the translational-rotational velocities of a microswimmer in the far-field domain as well as in the near-surface lubrication region.

Results reveal that hydrodynamic slippage mediates the competitive effects of the near-field hydrodynamic drag and propulsive forces. Consequently translational and rotational diffusion are both altered, sometimes even showing changes in sign, in contrast to the previous theoretical model of a far-field characteristics of force-dipole swimmer \citep{Lopez2014}. The pattern and intensity of the slip-induced changes in the swimming kinematics are critically dependent on the squirming modes and the distance of the microswimmer from the wall.

In comparison to the case of a no-slip wall, near-wall slip reduces the critical value of the dimensionless strength of the second squirming mode required to exert a sufficient hydrodynamic torque enabling a transformation of an escaping microswimmer trajectory to a wall-trapping one. Thus wall slip triggers a robust trapping nature of near-surface swimming states. Interestingly, depending upon the launching orientation angle and the strength of the swimming gait the slip-induced trapping can become either a periodic oscillation or a damped amplitude oscillations with a final fixed height and orientation.  However, in contrast to pullers, for which the critical slip length non-monotonically depends on the initial orientation angle, the critical slip length for pushers is independent of the initial orientation angle. In addition, the maximum attained height, the average height of the periodic oscillations and the final stable height, all get decreased with enhanced slip length.

We have identified that neutral swimmers do not show any tendency to get entrapped near a slippery wall; however, their detention time faces a significant enhancement with increasing wall slip and their minimum wall separation distance gets reduced before the wall-escape takes place. Additionally, as the slip length increases, the critical release orientation of the swimmer director for the transition from scattering to wall-collision, becomes more pointed away from the wall and finally reaches an asymptotic value.

 The present results may turn out to be elemental in providing a theoretical understanding of the complex behaviour of the natural microswimmers near a confinement boundary, either in a biophysical environment or those of the artificial swimmers in a controlled lab-on-a-chip device, where the substrate has hydrophobic surface property or there are wall-depleted zones in a bacterial polymeric solution.
Beyond the presently adopted spherical squirmer model of microorganisms, the incorporation of higher order squirming modes \citep{Pak2014} will be an interesting extension of the present work. It may also be stimulating research directions to additionally inspect the various aspects of micro-swimming such as elongated shape of microorganisms \citep{Ishimoto2013,Shum2010}, direct flagellar contact dynamics \citep{Kantsler2013} or the existence of thermal noise \citep{Li2009,Drescher2011,Schaar2015}. 

\section*{Acknowledgement}
S.C. acknowledges Department of Science and Technology, Government of India, for Sir J. C. Bose National Fellowship.
\begin{appendices}
	\renewcommand{\thesection}{Appendix \Alph{section}}
	\renewcommand{\thesubsection}{\thesection.\arabic{subsection}}
	\renewcommand\thefigure{A-\arabic{figure}}    
	\setcounter{figure}{0} 
	\renewcommand{\theequation}{A-\arabic{equation}}
	\setcounter{equation}{0}  

	\section{Details of the solution procedure}
	\label{sec:sol_detail}
 Following \cite{Lee1980} the expressions of the pressure and velocity field of the fluid in the cylindrical coordinates $ (u,v,w) $ in terms of eigenfunctions in the Bispherical coordinates, are given by
	\begin{equation}\label{key}
	p=\sum_{m=0}^{\infty} p_m (\xi,\eta) \cos(m\phi + \alpha_m),
	\end{equation}
	\begin{equation}\label{key}
	p_m=\frac{1}{c} \sqrt{(\cosh(\xi)-\zeta)} \sum_{n=m}^{\infty} \left[A_n^m \sinh(\beta_n \xi)+B_n^m \cosh(\beta_n \xi)  \right] P_n^m(\zeta).
	\end{equation}
	
	\begin{equation}\label{key}
	u=\frac{\rho p}{2} + u_0 \cos(\alpha_0) +\frac{1}{2} \sum_{m=1}^{\infty} (\gamma_m+\xi_m) \cos(m\phi +\alpha_m),
	\end{equation}
	\begin{equation}\label{key}
	v=v_0 \sin(\alpha_0) + \sum_{m=1}^{\infty} (\gamma_m-\xi_m) \sin(m\phi +\alpha_m),
	\end{equation}
	\begin{equation}\label{key}
	w=\frac{z p}{2} + \sum_{m=0}^{\infty} w_m \cos(m\phi +\alpha_m).
	\end{equation}
Here $ P_n^m $ is the associated Legendre polynomial of the first kind, $ \zeta=\cos(\eta)$ and $\beta_n=n+1/2$. Also 
\begin{equation}\label{key}
u_0=\sqrt{\cosh(\xi)-\zeta} \sum_{n=1}^{\infty} \left[E_n^0 \sin(\beta_n \xi) + F_n^0 \cosh(\beta_n \xi) \right] P_n^1(\zeta),
\end{equation}
\begin{equation}\label{key}
v_0=\sqrt{\cosh(\xi)-\zeta} \sum_{n=1}^{\infty} \left[G_n^0 \sin(\beta_n \xi) + H_n^0 \cosh(\beta_n \xi) \right] P_n^1(\zeta),
\end{equation}
\begin{equation}\label{key}
\gamma_m=\sqrt{\cosh(\xi)-\zeta} \sum_{n=m+1}^{\infty} \left[E_n^m \sin(\beta_n \xi) + F_n^m \cosh(\beta_n \xi) \right] P_n^{m+1}(\zeta),
\end{equation}	
\begin{equation}\label{key}
\chi_m=\sqrt{\cosh(\xi)-\zeta} \sum_{n=m-1}^{\infty} \left[G_n^m \sin(\beta_n \xi) + H_n^m \cosh(\beta_n \xi) \right] P_n^{m-1}(\zeta),
\end{equation}

\begin{equation}\label{key}
\text{and\qquad}w_m=\sqrt{\cosh(\xi)-\zeta} \sum_{n=m}^{\infty} \left[C_n^m \sin(\beta_n \xi)  \right] P_n^{m}(\zeta).
\end{equation}
The seven unknown constants ($ A_n^m,B_n^m, C_n^m, E^m_n, F_n^m, G_n^m, H^n_m $) appearing in the above expressions are obtained by satifying the boundary conditions at the swimmer surface \eqref{eq:BC_swimmer}, the Navier slip condition at the plane wall (\eqref{eq:simplify_slip}) and the continuity equation. 

In order to apply the swimmer surface boundary condition (\eqref{eq:BC_swimmer}), the surface velocity components are expanded in terms of bispherical eigenfunctions as given below
\begin{equation}\label{key}
u_s= \sum_{m} u_s^m (\xi,\eta) \cos(m\phi+\alpha_m),
\end{equation}
  \begin{equation}\label{key}
 v_s= \sum_{m} v_s^m (\xi,\eta) \sin(m\phi+\alpha_m),
 \end{equation}
  \begin{equation}\label{key}
  w_s= \sum_{m} w_s^m (\xi,\eta) \cos(m\phi+\alpha_m),
 \end{equation}
where for $ m=0 $,
\begin{subequations} \label{eq:uu}
\begin{equation}
u_s^0=\sqrt{\cosh(\xi_0)-\zeta} \sum X_n^0 (\xi) P_n^1(\zeta)
\end{equation}    
\begin{equation}
v_s^0=\sqrt{\cosh(\xi_0)-\zeta} \sum Y_n^0 (\xi) P_n^1(\zeta)
\end{equation}
\end{subequations}
for {$ m \ge 1 $},
\begin{subequations} \label{eq:X_Y_Z_0}
	\begin{equation}
	u_s^m+v_s^m=\sqrt{\cosh(\xi_0)-\zeta} \sum X_n^m (\xi) P_n^{m+1}(\zeta)
	\end{equation}    
	\begin{equation}
	u_s^m-v_s^m=\sqrt{\cosh(\xi_0)-\zeta} \sum Y_n^m (\xi) P_n^{m-1}(\zeta)
	\end{equation}
\end{subequations}
for all $ m $
\begin{equation}\label{eq:X_Y_Z_m}
w_s^m=\sqrt{\cosh(\xi_0)-\zeta} \sum Z_n^m (\xi) P_n^{m}(\zeta).
\end{equation}
The constants $ X_n^m,Y_n^m $ and $ Z_n^m $ are to be determined by using the boundary condition on the swimmer surface. Now making the use of the orthogonality of the associated Legendre polynomials, we obtain an infinite set of linear algebraic equations involving the unknown constants. Since the values of these constants decay with increasing values of $ n $, we truncate the algebraic system of equations for a large number of terms $ N $ so that the error in evaluating these constants ($ A_n^m,B_n^m, C_n^m, E^m_n, F_n^m, G_n^m, H^n_m $) as well as the swimmer velocity components $ (V_x,V_z, \Omega_y) $ between $ N $ and $ N+1 $ steps become $ \le 10^{-6} $. The system of equations has a banded matrix structure $ (7N\times 7N) $ and was solved using a numerical scheme. Similar to the previous works related to a passive or active sphere moving near a no-slip wall, we also find that the decay of these constants become very slow as the swimmer comes close to the plane wall which calls for a large number of terms to be retained to reach the desired accuracy \citep{Lee1980,Yazdi2017}. Adding to this, the increased value of the slip length at the plane wall turns out to be another hurdle to obtain an uniform accuracy throughout the calculations \citep{Loussaief2015,Kezirian1992}. Thus, for extreme cases when the swimmer is very close to wall ( e.g. $ h < 1.05$) and at the same time the wall slip length is very high (e.g. $l_{S}>5$), we work with an accuracy of $10^{-4}$ to save the computational cost. It has been verified that this does not cause any noticeable change in the results presented in this work.

Due to the linearity of Stokes equation, the flow field generated by a passively moving particle motion can be obtained by superposing the individual flow fields due to fundamental modes of the associated kinematics. Along similar lines, in the present problem of self propulsion, the flow fields due to the motion of a spherical particle with $V_x, V_z,\Omega_y$ in an otherwise quiescent fluid and that due to tangential slip velocity specified by the squirming modes $(B_1, B_2)$ for the case of a stationary sphere, each with a Navier slip condition at the plane wall, are sufficient to fully characterize the problem.  In the subsequent discussions, we denote each of these flow  fields as `fundamental problems'.
It is to be noted that the constants $ X_n^m,Y_n^m $ and $ Z_n^m $ in \eqref{eq:X_Y_Z_0} and \eqref{eq:X_Y_Z_m} come from the boundary condition at the microswimmer surface and thus remain unaffected by the slip at the plane wall. To avoid repetitiveness, we refer the reader to the earlier works \citep{Lee1980,Shaik2017} where these constants are provided for all the fundamental problems involved in the present study. The only difference arises from the fact that they have considered the body to be below the $ \xi=0 $ surface while we have taken the opposite configuration.
The flow fields due to a passive sphere moving near a slippery surface, having  translational and rotational velocity components parallel to the surface (e.g. $ V_x, \text{\,and\,}\Omega_y$), has been obtained  previously \citep{Kezirian1992,Loussaief2015} using a similar method as described above. However, the problem of a sphere moving normal to a slippery wall has only been solved using the streamfunction approach \citep{Goren1973}. The present authors have solved this problem using the direct solution of Stokes equation as discussed before. In addition, the flow problem due to the tangential squirming modes on the surface of a stationary sphere adjacent to a plane wall with fluid slip, has been solved for the first time. 

\section{Validation of the present numerical calculations}
	\label{sec:validation}
The solutions obtained from the numeric codes employed in the present study to obtain the full solution of the Stokes equation were first validated with various earlier works regarding a spherical particle motion near a no slip surface \citep{ONeill1964,Brenner1961} as well as near a slippery surface \citep{Kezirian1992,Loussaief2015,Goren1973}. Subsequently the force and torque values obtained from the full Stokes equation solution are rechecked with the aforementioned Reciprocal Theorem approach (section~\ref{ssec:reciprocal}).

 \begin{figure}[!htb]
	\centering
	\includegraphics[width=0.7\textwidth]{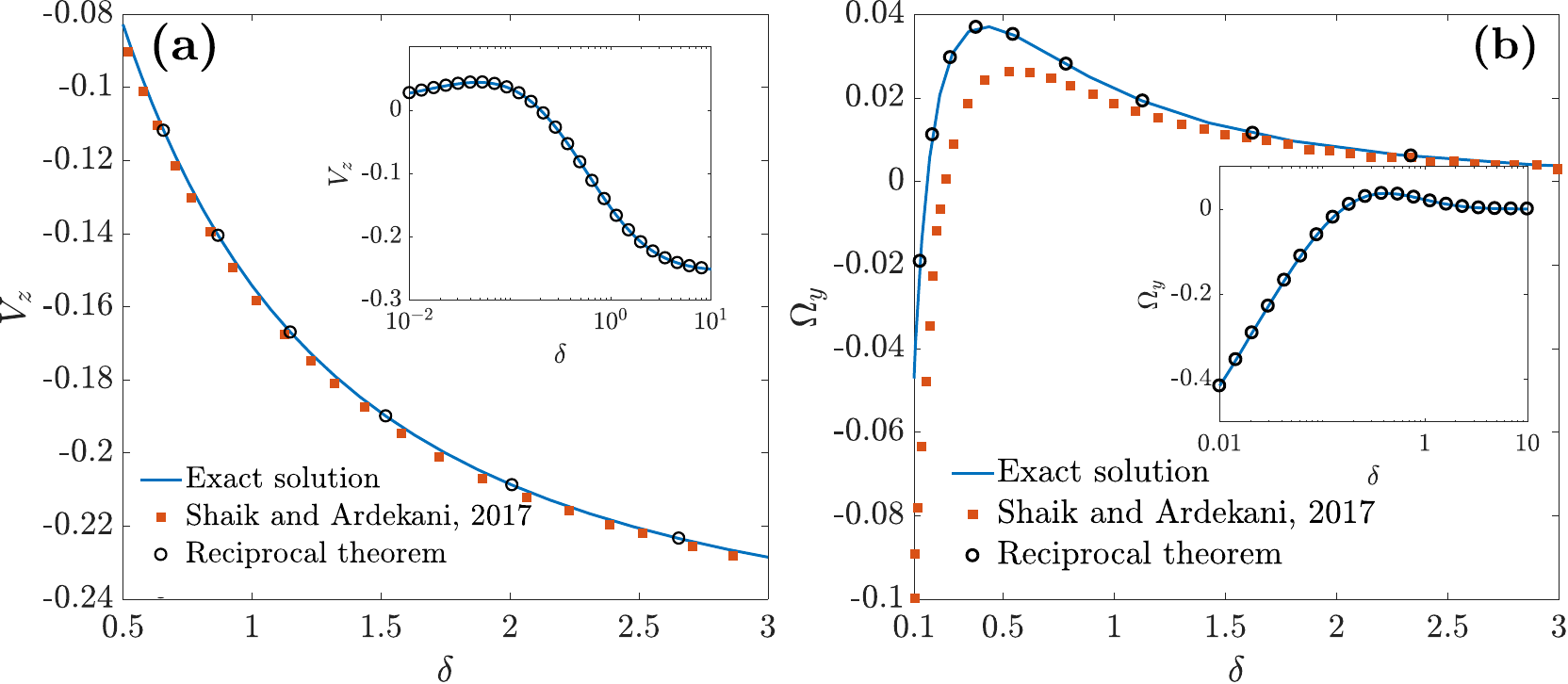}
	\vspace{1.5ex}
		\caption{Variation of the squirmer vertical velocity component and rotational velocity with the distance from the wall $(\delta)$. In the inset of each figure the log-scale variations are also highlighted. The parameters are: $B_1=1,\, \beta= 4$ and $\theta = \pi/8. $} 
	\label{fig:validation}
\end{figure}

In figures~\ref{fig:validation}(a),(b) we compare the no-slip results of vertical component of velocity and rotational velocity with the calculations of the Reciprocal theorem as well as with the previously reported results of \cite{Shaik2017}. The reciprocal theorem turns out to be matching almost exactly with the present exact solutions. The slight disagreement of the current results with those of \cite{Shaik2017} is due to the fact that  their results were obtained for a squirmer approaching a two fluid interface. We have taken their results corresponding to a high value of the viscosity ratio ($\lambda=10$), which only approximately resembles the characteristics of a no slip wall \citep{Lee1980}, while in an ideal case $ \lambda \to \infty $ is required to recover the results near a no-slip solid wall.

 \begin{figure}[!htb]
	\centering
	\includegraphics[width=0.8\textwidth]{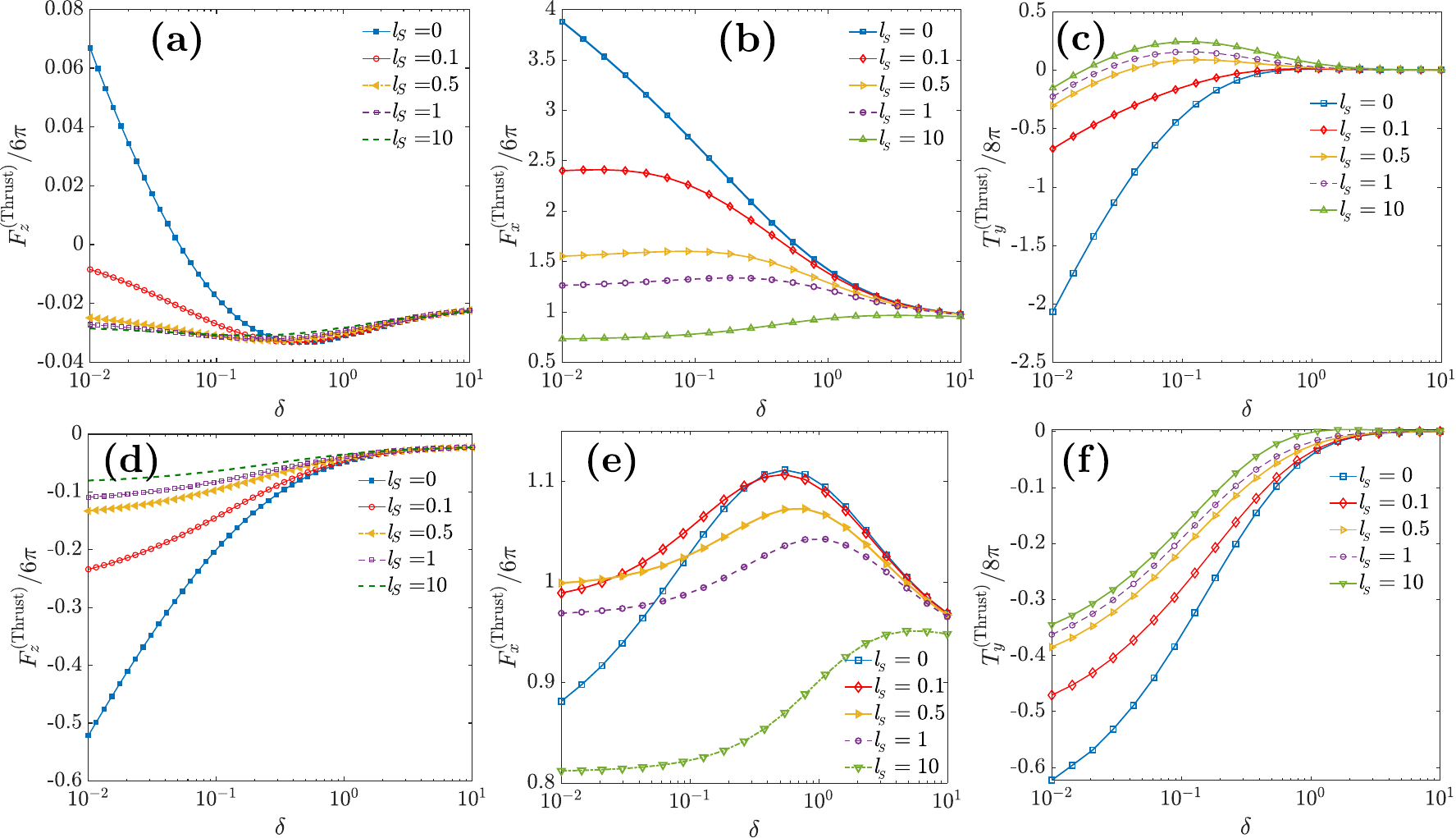}
	\vspace{1.5ex}
	\caption{Variation of thrust force and torque components $ (F_z^\text{(Thrust)}, F_x^\text{(Thrust)}, L_y^\text{(Thrust)}) $ with wall separation distance $ (\delta) $ for different slip lengths $ (l_{\!_S}) $. The parameters correspond to those of figure~\ref{fig:vel}.} 
	\label{fig:prop_force}
\end{figure}

 \begin{figure}[!htb]
	\centering
	\includegraphics[width=0.4\textwidth]{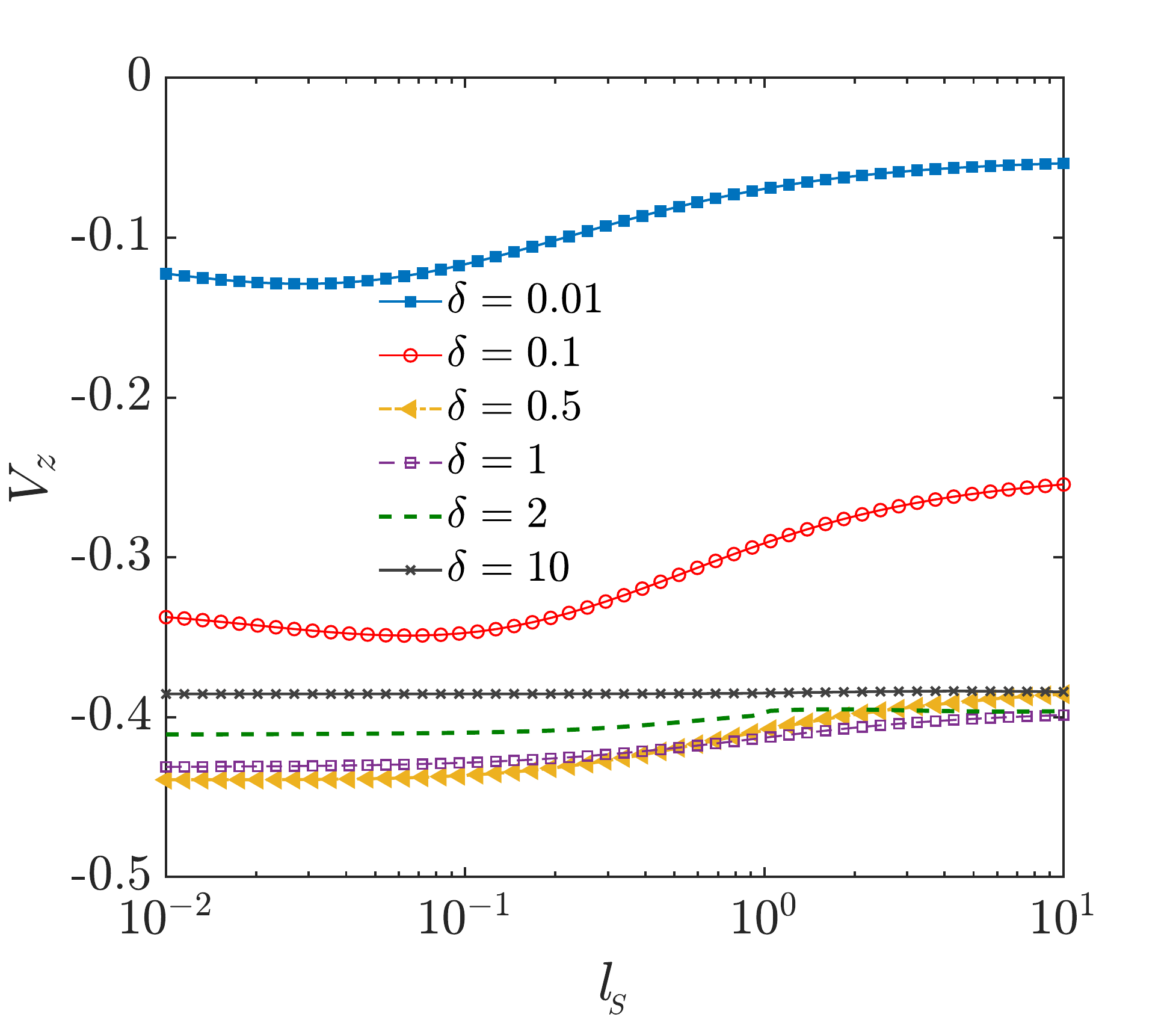}
	\vspace{1.5ex}
	\caption{Variation of wall normal velocity $ V_z $ of a pusher swimmer $( \beta=-4) $ with slip length for various wall separation distances $(\delta) $. Here microswimmer orientation, $\theta=\pi/8.$} 
	\label{fig:pusher_vz_suplle}
\end{figure}

\begin{figure}[!htb]
	\centering
	\includegraphics[width=0.7\textwidth]{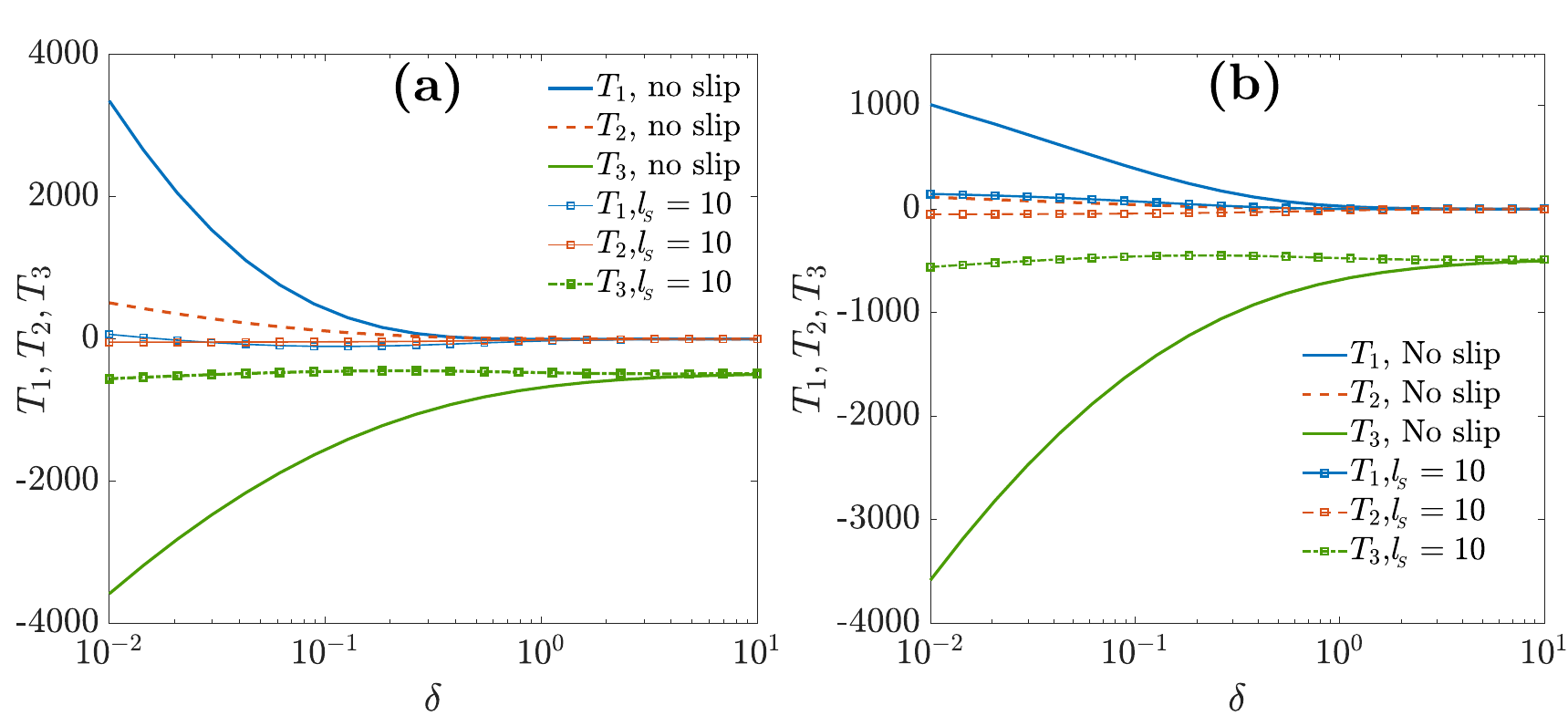}
	\vspace{1.5ex}
\caption{Three different terms $ (T_1,T_2,T_3) $ controlling the effective rotation rate of a (a) puller and (b)  pusher, as shown in \eqref{eq:Vx_Wy_coupled}(b). The parameters are same as figure~\ref{fig:vel}.}
\label{fig:Torque_3_terms}
\end{figure}

\begin{figure}[!htb]
	\centering
	\includegraphics[width=0.4\textwidth]{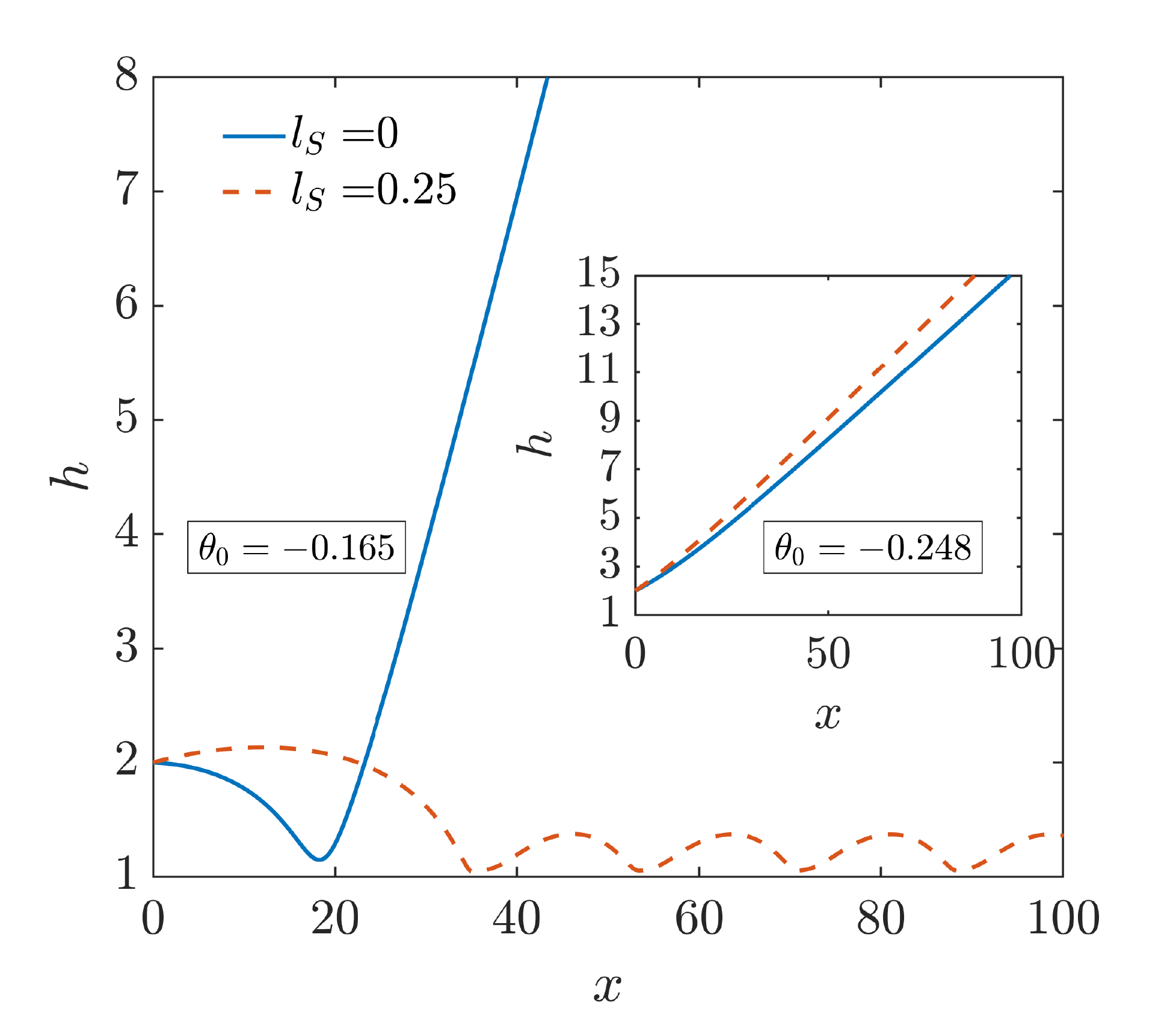}
	\vspace{1.5ex}
	\caption{Transition of swimming states for a pusher $(\beta=-5)$ near a  no-slip and slippery wall with $\theta_0=-0.165$. In the inset, the escaping states corresponding a more negative angle, $ \theta_0=-0.248$ are also shown.  } 
	\label{fig:Trajectory_beta_minus_5_TH0_minus_0p165_0p_248}
\end{figure}

 \begin{figure}[!htb]
	\centering
	\includegraphics[width=0.4\textwidth]{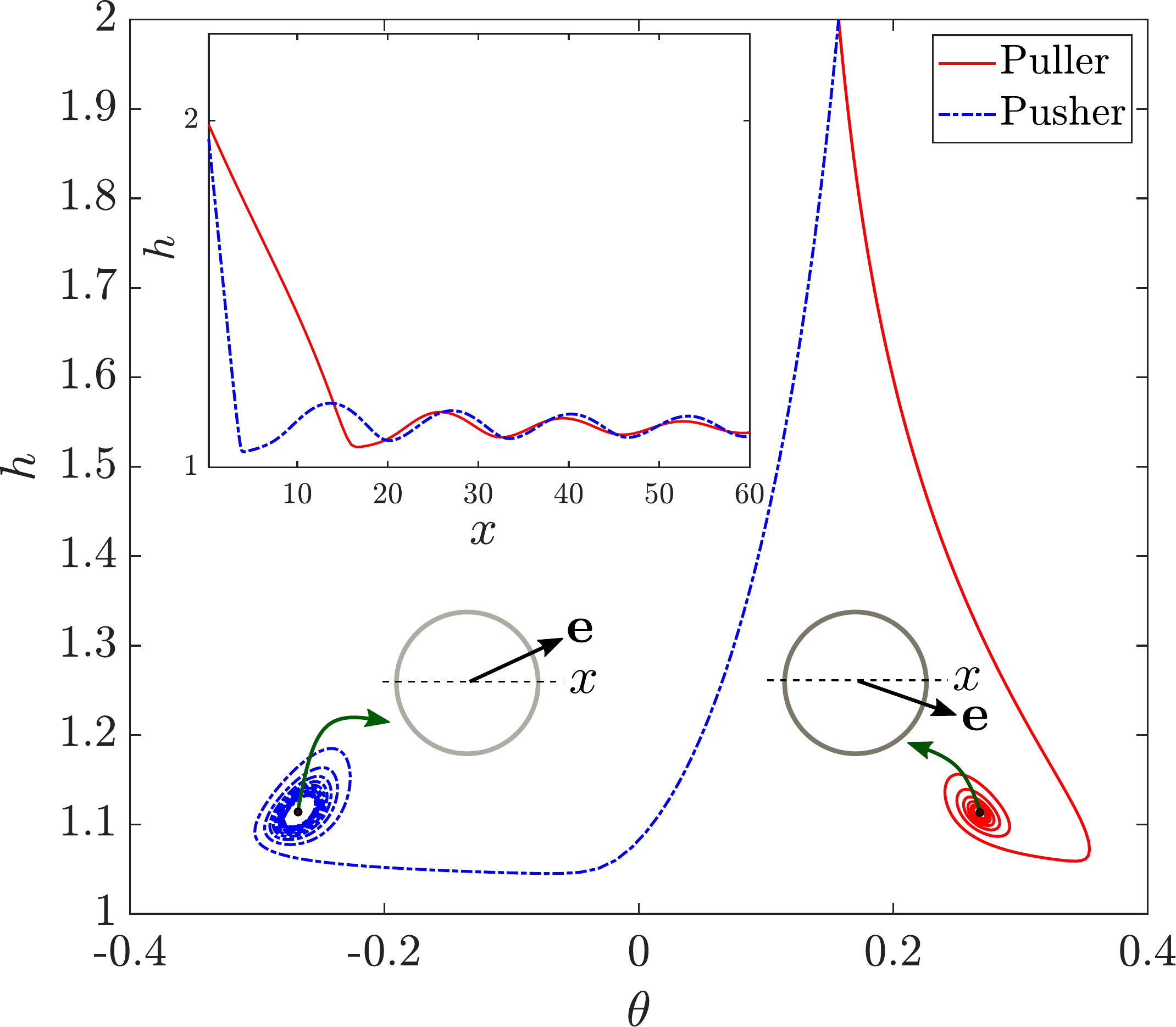}
	\vspace{1.5ex}
	\caption{Phase space trajectory comparison for puller and pusher microswimmers. The parameters are $ \delta_0=1,\theta_0=0.05\pi, l_{\!_S}=0.55$ and $|\beta|=4.$ In the inset the trajectories are compared in the $ (x,h) $ plane. Swimmer orientations during the final steady state swimming are also described schematically.} 
	\label{fig:Angle_comparison}
\end{figure}

\begin{figure}[!htb]
	\centering
	\includegraphics[width=0.7\textwidth]{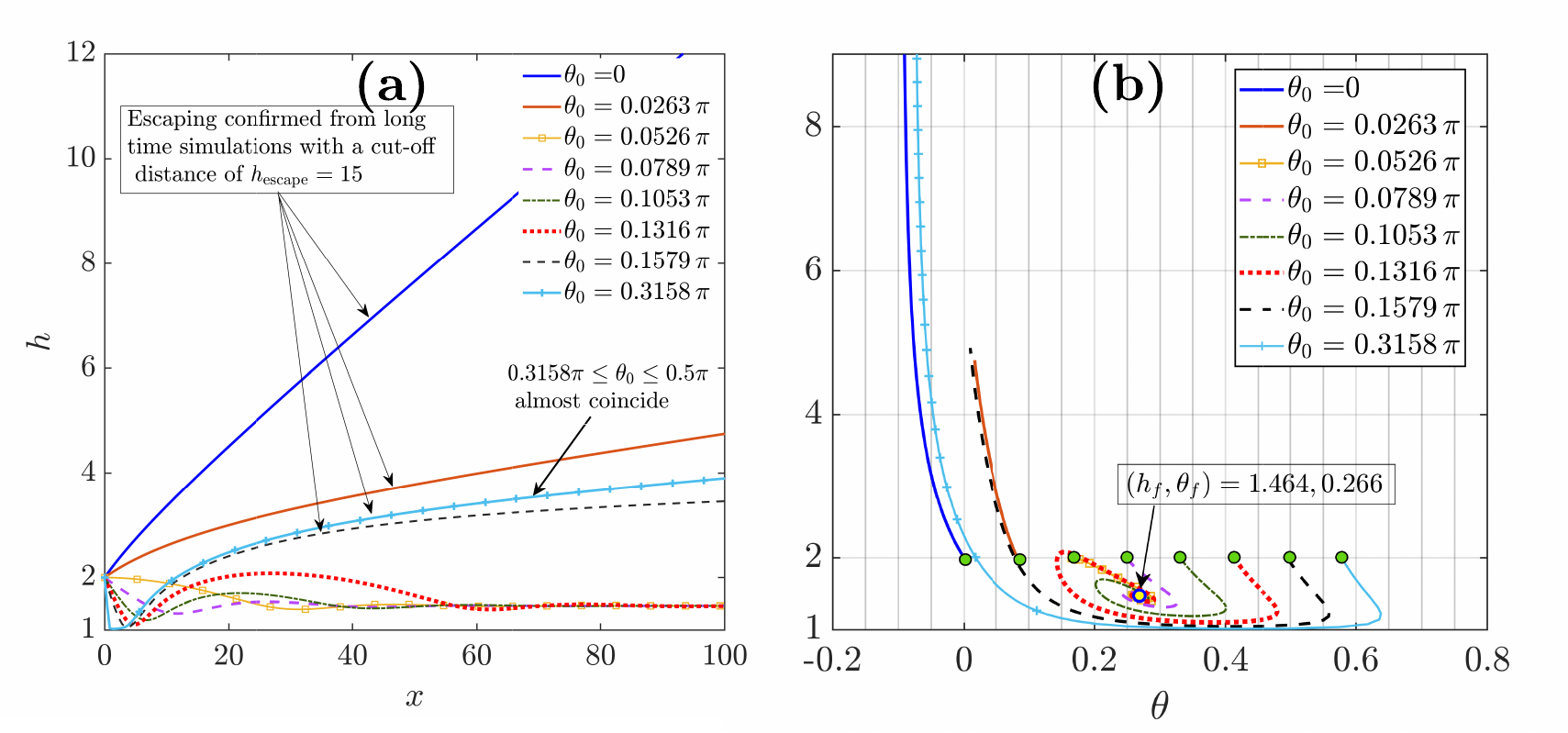}
	\vspace{1.5ex}
	\caption{Trajectory of a puller squirmer with $ \beta=5 $ and a wall slip length $ l_{\!_S}=0.03$ for various initial orientations $ (\theta_0).$ Subfigure: (a) Trajectory in the $(x,h)$ plane and (b) Phase plane dynamics,  $h$ vs. $\theta $. In (b), the green circles represent the initial states while the final fixed point for the trapping instances is shown with a circle.}
\label{fig:T_R_beta_5_Lambda_0p029764_non_monotonic_th0}
\end{figure}
\begin{figure}[!htb]
	\centering
	\includegraphics[width=0.4\textwidth]{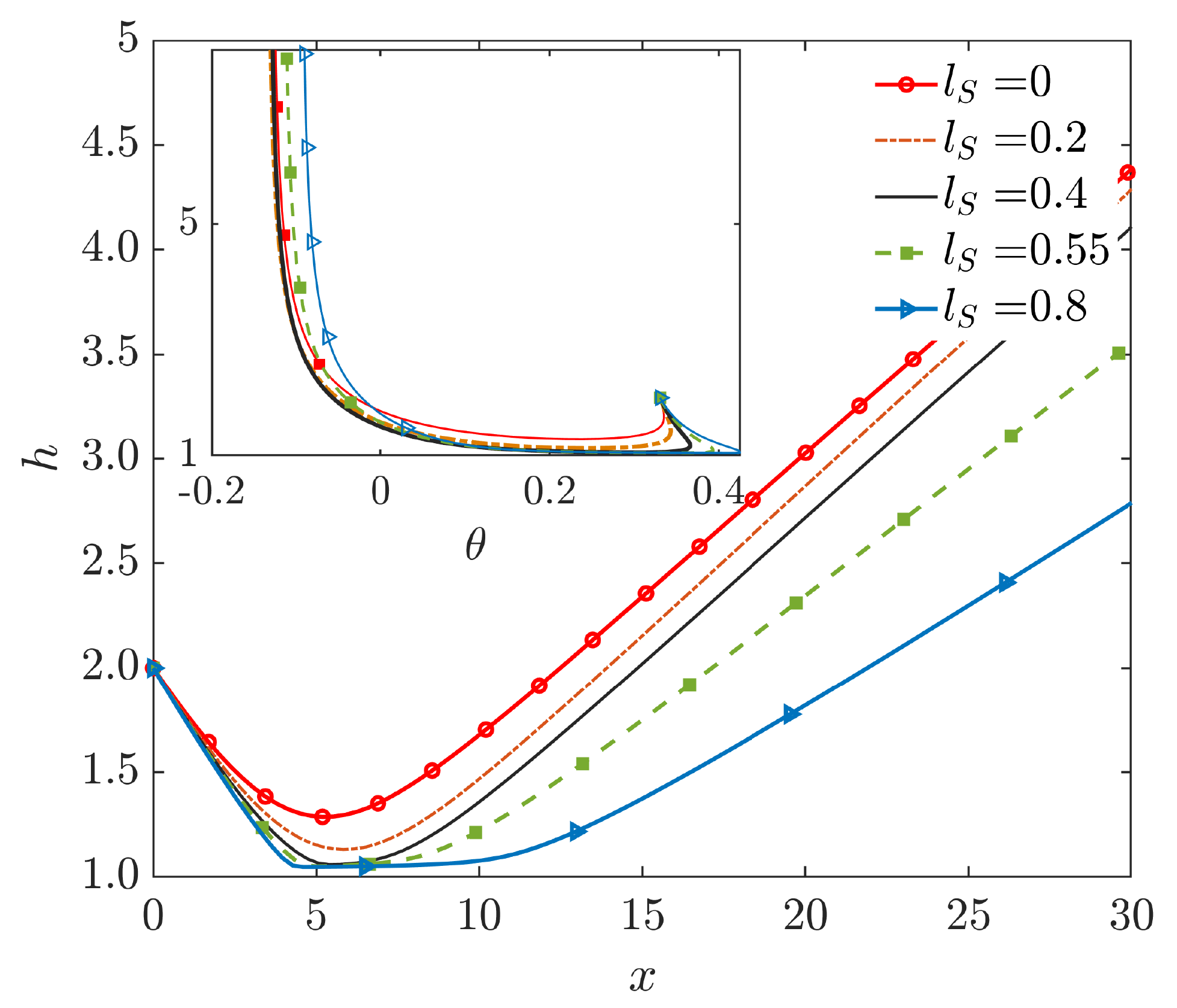}
	\vspace{1.5ex}
	\caption{Increase in detention time for a puller swimmer with $ \beta= 3, \theta_0 =0.1 \pi. $ } 
	\label{fig:Detention_time_beta_3_vary_ls_th0_0p1pi_T_R}
\end{figure}
\FloatBarrier

\end{appendices}


\bibliographystyle{jfm2}

\end{document}